\documentclass[11pt]{article}
\usepackage{amsfonts}
\usepackage{latexsym}
\usepackage{amssymb}
\usepackage{amscd}
\usepackage{amsmath}

\newlength{\dinwidth}
\newlength{\dinmargin}
\def\beq{\begin{equation}}
\def\eeq{\end{equation}}
\def\ben{\begin{displaymath}}
\def\een{\end{displaymath}}
\def\baa{\begin{eqnarray}}
\def\eaa{\end{eqnarray}}

\def\ba{\begin{array}}
\def\ea{\end{array}}
\makeatletter
\@addtoreset{equation}{section}
\makeatother

\def\lt{\tilde{\l}}
\def\gt{\tilde{\g}}
\def\eta{g}
\def\tauh{\tau}
\def\e{\epsilon}
\def\G{l}
\def\phi{\varphi}

\def\a{\alpha}
\def\g{\gamma}

\def\b{\beta}

\def\ka{\kappa}
\def\l{\lambda}

\def\t{\tau}

\def\O{\Omega}
\def\lp{x}
\def\hi{\chi}

\def\tro{\stackrel{1}{\tr}}
\def\trd{\stackrel{2}{\tr}}

\def\xibar{\bar{\xi}}
\def\t{\hat{\t}}

\def\C{{\mathbb C}}
\def\R{{\mathbb R}}
\def\CP1{{\mathbb C}P^1}

\def\la{\label}

\def\f{\frac}
\def\L{{\cal L}}
\def\p{\partial}

\def\tr{{\rm tr}}
\def\log{\ln}

\def\la{\label}

\def\f{\frac}
\def\p{\partial}
\def\res{{\rm res}}

\def\det{{\rm det}}

\def\fu{f}
\def\qb{{\bf q}}

\newtheorem{definition}{Definition}
\newtheorem{remark}{Remark}
\newtheorem{theorem}{Theorem}
\newtheorem{corollary}{Corollary}
\newtheorem{lemma}{Lemma}
\newtheorem{proposition}{Proposition}

\begin{document}

\title{A new hierarchy of integrable systems associated to  Hurwitz spaces}
%\shorttitle{}
\author{{A. Kokotov\footnote{e-mail: alexey@mathstat.concordia.ca}$\;$ and
D. Korotkin\footnote{e-mail: korotkin@mathstat.concordia.ca}}}
%\Names{}
%\Email{korotkin@mathstat.concordia.ca, alexey@mathstat.concordia.ca}
\maketitle

\begin{center}
Department of Mathematics and Statistics, Concordia University\\
7141 Sherbrooke West, Montreal H4B 1R6, Quebec,  Canada
\end{center}

\vskip0.5cm
{\bf Abstract.}
In this paper
we introduce a new class of integrable systems, naturally associated to
Hurwitz spaces (spaces of meromorphic functions over Riemann
surfaces). The critical values of the meromorphic functions  play the role
of ``times". Our systems give a natural generalization of the Ernst
equation; in genus zero they realize  the
scheme of deformation of integrable systems proposed by Burtsev,
Mikhailov and Zakharov. 
  We show that any solution of these systems in rank 1
defines a flat diagonal metric  (Darboux-Egoroff metric) together with
a class of corresponding systems of hydrodynamic type and their solutions.

\vskip0.8cm
\section{Introduction}

Deformations of Riemann surfaces appear in different aspects of the theory
of integrable systems. We  mention the theory of systems of hydrodynamic type
(in particular, the theory of Whitham deformations of integrable systems) \cite{Krich1,Dubr1}, theory of
algebro-geometric solutions of equations with variable spectral
parameter \cite{Koro88}
(the main representative of this class is  the Ernst equation from
general relativity) and the theory of Frobenius
manifolds \cite{Dubr94,Dubr98}.

The Ernst equation
\ben
\left((\xi-\xibar)G_{\xi} G^{-1}\right)_{\xibar}+
\left((\xi-\xibar)G_{\xibar} G^{-1}\right)_{\xi}=0\;,
\een
where $G(\xi,\bar{\xi})$ is a matrix-valued function (the  stationary
axially symmetric Einstein equations are equivalent to this equation
if $G$ is  a $2\times 2$
matrix with some additional symmetries) is the compatibility
condition of the following linear system ($U-V$ pair) \cite{BelZak78,Mais78}:
\begin{equation}
\f{\p\Psi}{\p\xi}=\f{G_\xi G^{-1}}{1-\nu}\Psi \;,\hskip1.0cm
\f{\p\Psi}{\p\bar{\xi}}=\f{G_{\bar{\xi}} G^{-1}}{1+\nu}\Psi \;,
\la{Ernstintr}
\end{equation}
where $\nu$ is the following function of spectral parameter ${\l}\in\C$
and variables
$(\xi,\bar{\xi})$:
\ben
\nu=\f{2}{\xi-\bar{\xi}}\left\{{\l}-\f{\xi+\bar{\xi}}{2}+\sqrt{({\l}-\xi)({\l}-\bar{\xi})}\right\}\;.
\een
Function $\nu({\l})$ is nothing but the uniformization map of the genus
zero Riemann surface which is the twofold covering of ${\l}$-plane with
two branch points at ${\l}=\xi$ and ${\l}=\bar{\xi}$. The map ${\l}(\nu)$ is
a rational map $\CP1\to\CP1$ of degree two with critical values $\xi$
and $\bar{\xi}$.

If the matrix dimension of $G$ is $1$, we can introduce the function
$f=\log G$ and the Ernst equation turns into the Euler-Darboux equation:
\begin{equation}
\f{\p^2 f}{\p\xi\p\bar{\xi}}+\f{1}{2(\bar{\xi}-\xi)}\f{\p f}{\p\xi}+\f{1}{2(\xi-\bar{\xi})}\f{\p f}{\p\bar\xi}=0\;.
\la{EDintr}
\end{equation}
The natural question is whether the Ernst equation is  an isolated example of an
integrable system related to  the space of rational maps or   it is possible to
define natural analogs of the Ernst equation which would correspond to
spaces of rational maps of arbitrary degree? More general, is it
possible to go beyond the spaces of rational maps, and define natural
analogs of the Ernst equation corresponding to general Hurwitz spaces $H_{g,N}$ of
meromorphic functions of degree $N$ on Riemann surfaces of genus $g$?
Is there any link between these higher analogs of Ernst equation and
existing theories of systems of hydrodynamic type \cite{Krich1,Dubr1}, Frobenius manifolds and
Darboux-Egoroff metrics corresponding to Hurwitz spaces \cite{Dubr94}?

If we assume that $\nu$ in (\ref{Ernstintr}) is a constant, then the
compatibility conditions of (\ref{Ernstintr}) leads to equation of
principal chiral model; therefore, according to terminology proposed
by Burtsev, Mikhailov and Zakharov \cite{BuMiZa}, it is natural to
call the Ernst equation the ``deformation'' of the principal chiral
model equation. In \cite{BuMiZa} it was studied the general problem of
deformation of a given integrable system which has $U-V$ pair, where
 matrices $U$ and $V$ are meromorphic functions of constant
spectral parameter $\nu$. If one allows $\nu$ to depend on
space variables $(x,y)$, then the zero curvature condition $U_y-V_x
+[U,V]=0$
implies a set of differential equations for ``variable spectral
parameter'' $\nu$ and poles of matrices $U$ and $V$. However, in
\cite{BuMiZa} no regular method was given to solve these differential
equations. 

The first goal of this paper is to fill this gap. Namely, we show that
the deformation scheme of \cite{BuMiZa} can be realized in terms of
the spaces of rational maps $H_{0,N}$ of any given degree $N$. In this way
we get a new hierarchy of non-autonomous nonlinear integrable systems.
We show how  solutions of the new  systems can be described
in terms of matrix Riemann-Hilbert problem and isomonodromic deformations.
If the matrix dimension equals $1$,
these non-linear  systems give rise to  systems of linear non-autonomous  second order partial
differential equations, generalizing
the Euler-Darboux equation.

Second, we  extend our framework  to  
construct a class of new integrable
systems starting from an arbitrary Hurwitz space $H_{g,N}$
(space of meromorphic functions of degree $N$ on Riemann surfaces of
genus $g$) for $g\geq 2$.

Third,  in rank 1 (by rank of the integrable system we
mean its matrix dimension) and any genus we observe a very close relationship
between  our systems, Darboux-Egoroff metrics, systems
of hydrodynamic type and Frobenius manifolds. Moreover, our general
formalism allows to give a simple description of systems of
hydrodynamic type (as well as their solutions) associated to Hurwitz
spaces $H_{g,N}$.

Let us describe our results in more details.
Each rational map $R(\g)$ of degree $N$ defines an $N$-fold
branch  covering ${\L}$ of $\CP1$; a point $P$ of ${\L}$ is a pair
$({\l},\g)$  such that 
\begin{equation}
{\l}=R(\g)\;.
\la{coveq}
\end{equation}
  We assume that  $R(\g)$ maps the infinity in $\g$-plane to
the infinity in ${\l}$-plane and $R(\g)={\l}+o(1)$ as $\g\to \infty$; then function $R(\g)$ has the form
\begin{equation}
R(\g)=\g+\sum_{m=1}^{N-1}\f{r_m}{\g-\mu_m}\;.
\la{ratint1}
\end{equation}
 Denote the critical
points of the map $R$ by $\g_1,\dots,\g_M$; we shall assume that all of them are simple; then  $M=2N-2$ according to
the Riemann-Hurwitz formula. 

The ramification  points of the covering ${\L}$ are denoted by  $P_m=({\l}_m,\g_m)$
(they are simple since oll critical points of $R(\gamma)$ are simple); their projections 
${\l}_m$  on ${\l}$-plane 
are called the branch points of the covering ${\L}$ (we adopt terminology of \cite{Fulton}); 
these are the values of the rational map
$R(\g)$ at its critical points: ${\l}_m=R(\g_m)$. In the sequel we shall assume that all ${\l}_m$ 
are different.

Introduce two functions on the covering ${\L}$: the function  $\pi$, which  projects ${\L}$ on 
${\l}$-plane: $\pi(P)={\l}$, and function $\nu$, which 
projects ${\L}$ to $\g$-plane: $\nu(P)=\g$. The map $\nu:{\L}\to \CP1$ is a one-to-one map; its 
inverse is nothing but uniformization map of
the covering ${\L}$. The maps  $\nu$ and $\pi$ are related as follows: $R(\nu(P))=\pi(P)$.                         

Due to our assumption about the behaviour of $R(\g)$ at
 infinity, in a neighbourhood of the infinite point
on some (we shall call it the first) sheet of ${\L}$ the map $\nu(P)$ behaves as follows:
$\nu(P)={\l}+o(1)$.

The structure of Riemann surface  on the branch covering ${\L}$ is
defined as follows: in a neighbourhood of any 
point where ${\L}$ is non-ramified we can consider  ${\l}$  as local parameter.
In a  neighbourhood of a ramification point $P_m$ the local coordinate is chosen to be 
$\lp_m=\sqrt{{\l}-{\l}_m}$.

The branch  covering ${\L}$ is completely defined by the branch points
${\l}_m$
and a representation of fundamental group
 $\pi_1(\CP1\setminus \{{\l}_1,\dots,{\l}_M\} )$ in permutation group
$S_N$. An element of permutation group describes permutation of sheets
 of the covering ${\L}$ if ${\l}$ encircles a given contour in $\CP1\setminus \{{\l}_1,\dots,{\l}_M\}$.
We shall consider local deformations of the covering ${\L}$ such that this representation is kept fixed. 
Then the branch points
${\l}_m$ can be considered as natural local coordinates on the space of
rational maps; they
will play the role of independent variables of our
systems.

Let us fix some point $P_0$ of ${\L}$ such that its projection on
${\l}$-plane ${\l}_0=\pi(P_0)$ is independent of all $\{{\l}_m\}$; let
$\g_0\equiv \nu(P_0)$.
Consider the following system of matrix linear differential equations
for an auxiliary matrix-valued $r\times r$ function $\Psi({\l},\{{\l}_m\})$:
\begin{equation}
\f{\p\Psi}{\p{\l}_m}=\f{\g_0-\g_m}{\nu(P)-\g_m} J_m \Psi\;.
\la{lsint}
\end{equation}
We  assume that the solution $\Psi$ of (\ref{lsint}) is normalized by the condition
\begin{equation}
\Psi({\l}=\infty)=I\;.
\la{normint}
\end{equation}
If we put in (\ref{lsint}) $P=P_0$, then 
the compatibility conditions of the  system (\ref{lsint}) imply that
all $J_m$ are logarithmic derivatives 
of a matrix-valued function $G(\{{\l}_m\})$: $J_m= G_{{\l}_m} G^{-1}$.
Besides that, the compatibility conditions of the  system (\ref{lsint}) imply the
following coupled system of non-autonomous non-linear matrix partial
differential equations  of second order:
\begin{equation}
\left((\g_0-\g_m) G_{{\l}_m} G^{-1}\right)_{{\l}_n}=\left((\g_0-\g_n) G_{{\l}_n} G^{-1}\right)_{{\l}_m}
\la{hieint}
\end{equation}
for all $m,n =1,\dots, M$.
We call  the matrix  dimension $r$ of the
matrix $G$  the rank of the system (\ref{hieint}).

For rational maps of degree two
the system (\ref{hieint}) coincides with the ordinary  Ernst equation  (\ref{Ernstintr}).
If the rank equals $1$, the systems (\ref{hieint})
gives rise to certain generalization of the  Euler-Darboux
equation (\ref{EDintr}).

We define  the tau-function of  these  systems by the following
system of compatible equations
\begin{equation}
\f{\p}{\p{\l}_m}\log\tau=\f{1}{2}\res\Big|_{P_m}\tr\left\{\f{(d_P\Psi\Psi^{-1})^2}{d{\l}}\right\}\;,
\la{tauint1}
\end{equation}
where $d_P\Psi=(\p\Psi/\p\nu(P)) d\nu(P)$; these equations fix $\tau$ up to an
arbitrary constant multiplier if we assume that $\tau$ is 
holomorphic function of $\{{\l}_m\}$.

Each  system (\ref{hieint}) possesses a subclass of ``isomonodromic'' solutions which can be built from
an arbitrary solution of the Schlesinger system in the same matrix dimension. For this isomonodromic sector
of  solutions we link  the tau-function to the Jimbo-Miwa
tau-function of the Schlesinger system. This link generalizes the formula
(found in \cite{KorNic96})
relating the so-called  ``conformal factor'' of the Ernst equation
with Jimbo-Miwa tau-function. We also show how to construct solutions
of the systems (\ref{hieint}) from solutions of
matrix Riemann-Hilbert problems on $\CP1$.

After developing the theory of the systems (\ref{hieint}), which are  related to the space
of rational maps $H_{0,N}$, we formulate a similar
construction based on an arbitrary Hurwitz space $H_{g,N}$.

Consider in detail the rank 1 case. Let the $N$-fold branched
covering ${\L}$ have genus $g$; then the number of branch points is equal
to $M=2g+2N-2$ (as before, we assume that all the branch points are simple
and have different projections on ${\l}$-plane). Again, the projections ${\l}_m$ of the branch points
$P_m$ on ${\l}$-plane can be used as local coordinates on $H_{g,N}$.
Let us introduce some basis of canonical cycles on ${\L}$. Denote by
$E(P,Q)$ the prime-form on ${\L}$, and by $B_m$ the normalized (all
$a$-periods vanish)  abelian
differential of the second kind with the unique pole of second order at $P_m$
with leading coefficient $1$.
Let us choose two arbitrary points $P_0$ and $Q_0$ on ${\L}$ such that their
projections $\pi(P_0)$ and $\pi(Q_0)$ on ${\l}$-plane are $\{{\l}_m\}$-independent.
Then the $H_{g,N}$ analog of the genus zero systems (\ref{hieint}) in rank 1 
looks as follows:
\begin{equation}
\f{\p^2 f}{\p{\l}_m\p{\l}_n}-\f{b(P_m,P_n)}{2}
\left\{\f{v_n}{v_m}\f{\p f}{\p{\l}_m}+
\f{v_m}{v_n}\f{\p f}{\p{\l}_n}\right\}=0\;,\hskip0.8cm m\neq n
\la{EDgenusint}
\end{equation}
where
\ben
b(P_m,P_n)= \frac{B(P,Q)}{dx_m(P) dx_n(Q)}\Big|_{P=P_m,\, Q=P_n}
\een
and $B(P,Q)=d_P d_Q\log E(P,Q)$ is the Begmann kernel;
$v_m =\int_{Q_0}^{P_0} B_m\;$.

Solutions of the system (\ref{EDgenusint}) can be constructed as
follows. 
Let $l$ be an arbitrary closed contour on ${\L}$ such that its
projection on the ${\l}$-plane $\pi(l)$ is independent of $\{{\l}_m\}$ and
$P_m\not\in l$ for any $m$. Let $h(P)$ be an arbitrary independent
of $\{{\l}_m\}$ H{\"o}lder-continuous
function on $l$. 
Then the function
\begin{equation}
f =\oint_{l} h(P) d_P\log\f{E(P_0,P)}{E(Q_0,P)} 
\la{solint00}
\end{equation}
satisfies the  system (\ref{EDgenusint}).

The systems (\ref{EDgenusint}) turn out to be in close
relationship with diagonal flat metrics  (Darboux-Egoroff
metrics). Namely, consider the diagonal metric
\begin{equation}
ds^2=\sum_{m=1}^M \eta_{mm} d{\l}_m^2\;,
\la{metricint}
\end{equation}
where
$\eta_{mm}=\p\log\tauh/\p{\l}_m$
and $\tauh(\{{\l}_m\})$ is the tau-function  corresponding to an
arbitrary solution of (\ref{EDgenusint}). The rotation coefficients of
this metric turn out to be given by   the Bergmann kernel:
\ben
\beta_{mn} = \f{1}{2}b(P_m,P_n)\;,
\een
i.e. they depend only on the  covering ${\L}$ and don't depend
on a particular solution of (\ref{EDgenusint}). In accordance with
the Rauch variational formulas \cite{Fay92}, which describe dependence
of the  Bergmann kernel on the branch points,  these coefficients
satisfy the equations
\ben
\f{\p \beta_{mn}}{\p{\l}_l}=\beta_{ml}\beta_{ln}
\een
for distinct $l,m,n$,  which, together with annihilation of all $\beta_{mn}$
by the operator $\sum_{k=1}^M\p/\p{\l}_k\;,\;$ guarantee flatness of the metric
(\ref{metricint}). 

The Darboux-Egoroff metrics are known to be closely related to integrable classes of systems of  hydrodynamic type, 
including, in particular,
Whitham equations which describe slow deformations of the Riemann surfaces arising in dispersionless limit of finite-gap solutions
of integrable systems. These systems are solvable via so-called generalized hodograph method \cite{Tsar90,Krich1}.
 Our present scheme gives a short and simple  formulation 
of the theory of the systems of hydrodynamic type corresponding to
Hurwitz spaces.

We notice that the present paper has some overlap with the paper
\cite{Krich3}, devoted to  isomonodromy deformations in higher genus,
 which appeared simultaneously with the first version of
this text \cite{KokKorpre}.

The paper is organized as follows. In section 2 we derive an auxiliary system of
differential equations which describe  dependence of the critical points
of the  rational map of the form (\ref{ratint1}) on the critical values. 
Then we show integrability of  the systems (\ref{hieint}), define  corresponding tau-function,
and
discuss the relationship of these systems to the
Riemann-Hilbert problem and the Schlesinger system. Here we also show
how the systems (\ref{hieint}) are related to
 deformations of integrable systems proposed in \cite{BuMiZa}.
In section 3 we define analogs of the systems (\ref{hieint}) related to
Hurwitz spaces $H_{g,N}$ in arbitrary genus $g\geq 2$, and discuss their properties.
In Sect.4 we  show that each   solution of  these systems in rank 1   defines a
Darboux-Egoroff metrics, and discuss related systems of hydrodynamic
type in our framework, together with their solutions.
Here we also outline the link between higher genus analogs of systems (\ref{hieint}) and
isomonodromic deformations on algebraic curves.
In section 5 we discuss  potential
directions of  future work.

\section{Non-autonomous integrable systems  related to spaces  of rational maps}

\subsection{Differential equations for critical points of rational maps}

Consider a rational map $R(\g)$ of degree $N$ of the form
(\ref{ratint1}). 
 If we choose the critical values  ${\l}_1,\dots,{\l}_M$ of the map
(\ref{ratint1}) as independent parameters, 
each critical point $\g_m$ becomes
a function of all $\{{\l}_m\}$; the study of these functions is the main subject of this subsection.
The function $\nu(P)$ depends on the variables
${\l}_1,\dots,{\l}_M$ as parameters. In the sequel we shall denote the
point of ${\L}$ which belongs to the $j$th sheet and has a projection
${\l}$ on the ${\l}$-plane by ${\l}^{(j)}$.
The map $\nu(P)$ has its only pole at the  point at infinity of some sheet of ${\L}$; we
enumerate the sheets of ${\L}$ in such a way that this sheet has number one;
therefore,
\begin{equation}
\nu(P)= {\l} +o(1)\ \ \ \ \  \text{as}\ \ P\to \infty^{(1)}.
\la{ginfty}
\end{equation}

\begin{theorem}
Function $\nu(P)$, considered locally as function of ${\l}$ and
depending on  branch points ${\l}_1,\dots,{\l}_M$ as
parameters, satisfies the following equations:
\begin{equation}
\f{\p\nu}{\p{\l}}=1+ \sum_{n=1}^M\f{\a_n}{\nu-\g_n}\;,\hskip1.0cm
\f{\p\nu}{\p{\l}_n} = -\f{\a_n}{\nu-\g_n}\;.
\la{glln}
\end{equation}
where  $\a_m$ are some functions of the branch points.
\end{theorem}

{\it Proof.}
Consider the local behavior of the function $\nu(P)$ near a branch point:
\begin{equation}
\nu(P) = \g_m + \ka_m \sqrt{{\l}-{\l}_m} + O({\l}-{\l}_m) \hskip0.5cm {\rm as}  \hskip0.5cm P\to P_m\;.
\la{gbranch}
\end{equation}
From (\ref{gbranch}) we conclude that
\ben
\f{\p\nu}{\p{\l}}= \f{\ka_m}{2\sqrt{{\l}-{\l}_m}}+ O(1)\;,\hskip1.0cm
\f{\p\nu}{\p{\l}_n}=-\delta_{mn} \f{\ka_m}{2\sqrt{{\l}-{\l}_m}} + O(1)\;,
\een
as $P\to P_m$.
By (\ref{gbranch}), we can rewrite these expansions using $\nu(P)$ as the global coordinate on ${\L}$:
\begin{equation}
\f{\p\nu}{\p{\l}}= \f{\ka_m^2}{2(\nu-\g_m)}+ O(1)\;,\hskip1.0cm
\f{\p\nu}{\p{\l}_n}=-\delta_{mn}\f{\ka_m^2}{2(\nu-\g_n)}+ O(1)\;,
\la{locder}
\end{equation}
where $\g_n = \nu(P_n)$.
Moreover, from (\ref{ginfty}) we conclude that ${\p\nu}/{\p{\l}}=1 + o(1)$ and  ${\p\nu}/{\p{\l}_n}=o(1)$ as $\nu(P)\to \infty$.
Therefore, ${\p\nu}/{\p{\l}}$ is a meromorphic function of $\nu$ with simple
poles at all the points $\g_n$ with residues $\ka_n^2/2$
and value $1$ at infinity. Analogously, the function ${\p\nu}/{\p{\l}_n}$
is a meromorphic function on $\CP1$ with simple pole at
${\l}_n$ and zero at infinity.
Therefore, we get equations (\ref{glln}) with
$\a_n = \ka_n^2/2$. 

$\Box$

\begin{corollary}
The critical points $\g_m$ of the rational function (\ref{ratint1}) and residues $\a_m$ from (\ref{glln}) depend as follows
on the critical values ${\l}_m$:
\begin{equation}
\f{\p\g_m}{\p{\l}_n}= \f{\a_n}{\g_n-\g_m}\;,\hskip0.6cm m\neq n\;;\hskip0.6cm 
\f{\p\g_m}{\p{\l}_m}= 1+\sum_{n=1,\,n\neq m}^M\f{\a_n}{\g_m-\g_n}\;,
\la{glmn}
\end{equation}
\begin{equation}
\f{\p\a_m}{\p{\l}_n}= \f{2\a_n\a_m }{(\g_n-\g_m)^2}\;,\hskip0.6cm m\neq n\;;\hskip0.6cm 
\f{\p\a_m}{\p{\l}_m}= -\sum_{n=1,\,n\neq m}^M\f{2\a_n\a_m }{(\g_n-\g_m)^2}
\la{almn}
\end{equation}
for all $m,n = 1,\dots,M$.
\end{corollary}

{\it Proof.} Equations (\ref{glmn}) and (\ref{almn}) follow from compatibility of equations (\ref{glln}).

$\Box$

Rational functions of the form (\ref{ratint1}) were introduced by Kupershmidt and Manin \cite{KupMan} in connection with
Benney systems. 
The fact that the  critical points of these functions  satisfy
equations (\ref{glmn}),(\ref{almn}), follows from the recent paper 
by Gibbons and Tsarev \cite{GibTsa99} \footnote{We thank E.Ferapontov,
who attracted our attention to this work}.
However, we did not find in existing literature the whole set of equations
(\ref{glln}), (\ref{glmn}), (\ref{almn})
associated to the functions  (\ref{ratint1}).

{\bf Two-fold coverings.}
For an illustration consider the case
$N=2$, when the  covering ${\L}$ has
 two sheets and two branch points ${\l}_1$, ${\l}_2$.
Then the rational function $R(\g)$ (\ref{ratint1}) can be explicitly
written in terms of
its critical values  ${\l}_1$ and ${\l}_2$: 
\begin{equation}
R(\g)=\g+\f{\left({\l}_1-{\l}_2\right)^2}{16(\g-\f{{\l}_1+{\l}_2}{2})}\;.
\end{equation}
The map $\nu(P)$ looks  as follows:
\begin{equation}
\nu(P)= \f{1}{2}\Big({\l}+\f{{\l}_1+{\l}_2}{2}+\sqrt{({\l}-{\l}_1)({\l}-{\l}_2)}\Big)\;.
\la{unifor}
\end{equation}
The critical points $\g_{1,2}$ and variables  $\a_{1,2}$ are given by:
\begin{equation}
\g_1\equiv\nu({\l}_1)=\f{3{\l}_1+{\l}_2}{4}\;,\hskip1.0cm \g_2\equiv\nu({\l}_2)=\f{{\l}_1+3{\l}_2}{4}\;,
\la{g12}
\end{equation}
\begin{equation}
\a_1= -\a_2 = \f{{\l}_1-{\l}_2}{8}\;.
\la{a12}
\end{equation}

\subsection{Spaces of rational maps and new hierarchy of
non-autonomous integrable systems}

Starting from 
  an arbitrary branch $N$-fold covering ${\L}$ of genus zero we can construct a
  hierarchy of  integrable
systems as follows.

Fix some point $P_0\in{\L}$ such that its projection ${\l}_0$ on $\CP1$ is independent of all $\{{\l}_m\}$, i.e.
$\g_0\equiv \nu(P_0)$ depends on $\{{\l}_m\}$ according to the equation
$$
R(\g_0({\l}_1,\dots,{\l}_M))={\l}_0\;.
$$
Consider the following system of  first order differential
equations for a $r\times r$ matrix-valued function  $\Psi(P,\{{\l}_m\})$:
\begin{equation}
\f{d\Psi}{d{\l}_m}=\f{\g_0-\g_m}{\nu(P)-\g_m}J_m \Psi\;,
\la{ls}
\end{equation}
where $J_m$ are $r\times r$ matrix-valued functions of $\{{\l}_m\}$.

As a corollary of compatibility conditions of the linear system
 (\ref{ls})  functions $J_m$
 can be expressed in terms of the single function
$G\equiv \Psi(P_0)$:
\begin{equation}
J_m=\f{\p G}{\p{\l}_m}G^{-1}\;;
\la{JmG}
\end{equation}
moreover, the function $G$ satisfies the following system  of non-linear partial differential equations:
\begin{equation}
\left((\g_0-\g_m) G_{{\l}_m} G^{-1}\right)_{{\l}_n}=\left((\g_0-\g_n) G_{{\l}_n} G^{-1}\right)_{{\l}_m}\;.
\la{hie1}
\end{equation}
(to derive (\ref{hie1}) from (\ref{ls}) one needs to make use of
equations (\ref{glln}),  (\ref{glmn}) and (\ref{almn}).

We write the ``total'' derivative of $\Psi$ with respect to ${\l}_m$ in
(\ref{ls}) to emphasize that in (\ref{ls})  we consider ${\l}$ and
${\l}_1,\dots,{\l}_m$ as independent variables. 

Alternatively, 
consider $\Psi(P)$ as a function of $\nu(P)$ and ${\l}_1,\dots,{\l}_M$.  Then, since
$\nu(P)$ is itself a  function of ${\l}$ and $\{{\l}_m\}$, we can
rewrite (\ref{ls}) using the chain rule
\ben
\f{d\Psi}{d{\l}_m}=\f{\p\Psi}{\p{\l}_m} +\f{\p\nu}{\p{\l}_m}\f{\p\Psi}{\p\nu}\;,
\een
where the notation $\p\Psi/\p{\l}_m$ means that $\nu(P)$ remains fixed.

Then, using (\ref{glln}) for $\p\nu/\p{\l}_m$, we rewrite  (\ref{ls}) as follows:
\begin{equation}
\f{\p\Psi(P)}{\p{\l}_m}
-\f{\a_m}{\nu(P)-\g_m}\f{\p\Psi(P)}{\p\nu(P)}=
\f{\g_0-\g_m}{\nu(P)-\g_m}J_m \Psi(P)\;.
\la{ls1}
\end{equation}

A simple calculation  making use of  equations (\ref{hie1}) and
 system (\ref{glmn}), (\ref{almn}), shows that if 
 $G(\{{\l}_m\})$ is a solution of non-linear system (\ref{hie1}), then the  1-form
\begin{equation}
\qb= \sum_{m=1}^M \f{(\g_0-\g_m)^2}{2\a_m}\tr\left(G_{{\l}_m} G^{-1}\right)^2 d{\l}_m
\la{1form}
\end{equation}
is  closed, $d\qb=0$.

The closedness of the 1-form $\qb$ implies  the existence of the
potential $\tauh$ , 
which can naturally be called the tau-function of the non-linear
system (\ref{hie1}).
\begin{definition}
The function $\tauh({\l}_1,\dots,{\l}_M)$, defined  by the following system of
equations:
\begin{equation}
\f{\p\log\tauh}{\p{\l}_m} =\f{(\g_0-\g_m)^2}{2\a_m}\tr\left(G_{{\l}_m} G^{-1}\right)^2
\la{tauhat}
\end{equation}
up to an arbitrary constant multiplier, is called the tau-function of
integrable system (\ref{hie1}).
\end{definition}

Using equations (\ref{hie1}) (\ref{glmn}), (\ref{almn}), we find that the
 second derivatives of the tau-function are given by the following expression:
\begin{equation}
\f{\p^2\log\tauh}{\p{\l}_m\p{\l}_n} = \f{(\g_0-\g_m)(\g_0-\g_n)}{2(\g_m-\g_n)^2}\tr \left\{G_{{\l}_m} G^{-1}G_{{\l}_n} G^{-1}\right\}
\la{tauhll}
\end{equation}
for $m\neq n$.

 Taking the residue of the  linear system (\ref{ls1}) at $P=P_m$, we
 find
\begin{equation}
\a_m\Psi_\g\Psi^{-1} (P_m) = (\g_0-\g_m)G_{{\l}_m} G^{-1}\;;
\la{psig}
\end{equation}
due to this relation the definition (\ref{tauhat}) allows an alternative
formulation:

{\bf Definition 1'} {\it The tau-function of the system (\ref{hie1}) is defined
by the following  equations:}
\begin{equation}
\f{\p\log\tauh}{\p{\l}_m}=\f{1}{2}\res|_{P_m}\left\{\f{\tr(d_P\Psi\,\Psi^{-1})^2}{d{\l}}\right\}\;,
\la{tauhat2}
\end{equation}
where  
\begin{equation}
d_P\Psi\,\Psi^{-1}\equiv \Psi_{\nu(P)}\Psi^{-1} d\nu(P)\;.
\la{qbdef}
\end{equation}
\vskip0.3cm

Let us  prove the equivalence of the two definitions of the tau-function.
Using $\nu(P)$ as a global coordinate on ${\L}$, we have:
\ben
d{\l}=\f{\p{\l}}{\p\nu} d\nu\;;
\een
therefore,
\ben
\f{\tr\{(d_P\Psi\,\Psi^{-1})^2 \}}{d{\l}} = \f{\p\nu}{\p{\l}}\tr\{ (\Psi_\nu\Psi^{-1})^2\} d\nu=
\left\{1+\sum_{m=1}^M\f{\a_m}{\nu-\g_m}\right\}\tr\{ (\Psi_\nu\Psi^{-1})^2\} d\nu\;,
\een
and (\ref{tauhat2}) coincides with (\ref{tauhat}).

\vskip0.3cm

{\bf The Ernst equation.} For $N=2$  the hierarchy
(\ref{hie1}) reduces to a single  equation. If one chooses point 
$P_0$
to coincide with $\infty^{(2)}$
[i.e. the  point of ${\L}$ where ${\l}=\infty$ and in a neighbourhood of
which
 $\sqrt{({\l}-{\l}_1)({\l}-{\l}_2)}=-{\l}+({\l}_1+{\l}_2)/2+o(1)\,$], we get
\ben
\g_0=\nu(P_0) = \f{{\l}_1+{\l}_2}{2}\;.
\een
Taking into account expressions (\ref{g12}), we have
\begin{equation}
\g_0-\g_1=\f{{\l}_2-{\l}_1}{4}\;,\hskip1.0cm \g_0-\g_2=\f{{\l}_1-{\l}_2}{4}\;.
\la{coeff12}
\end{equation}
If we now assume that ${\l}_1$ and ${\l}_2$ are conjugated to each other:
${\l}_1=\xi$, ${\l}_2=\xibar$, then equation
(\ref{hie1}) takes the following form:
\begin{equation}
\left((\xi-\xibar)G_{\xi} G^{-1}\right)_{\xibar}+
\left((\xi-\xibar)G_{\xibar} G^{-1}\right)_{\xi}=0\;.
\la{Ernst}
\end{equation}
This equation is called the Ernst equation; it is equivalent to vacuum
Einstein's equation for  stationary axially symmetric
spacetimes (the matrix $G$ in this case must be real, symmetric and
must have unit determinant).
In this case the linear system (\ref{ls})  is equivalent to the Lax representation of the Ernst equation found in 1978 by
Belinskii-Zakharov \cite{BelZak78} and Maison \cite{Mais78}.  We
notice that the Maison's Lax pair is  a
partial case of our linear system written in the form (\ref{ls});
whereas  the Belinskii-Zakharov linear system is a partial case of  (\ref{ls1}).

Due to expressions (\ref{a12}) for $\a_{1,2}$,
the definition of the tau-function  $\tauh$ can be written down  as follows:
\begin{equation}
\f{\p\log\tauh}{\p\xi}=\f{\xi-\xibar}{4}\tr\left(G_{\xi} G^{-1}\right)^2\;,\hskip1.0cm
\f{\p\log\tauh}{\p\xibar}=\f{\xibar-\xi}{4}\tr\left(G_{\xibar} G^{-1}\right)^2\;.
\la{tauernst}
\end{equation}
Formula (\ref{tauernst}) coincides with the definition of the so-called conformal factor - one of the metric coefficients
which correspond to a  given solution of the Ernst equation.

\subsection{New integrable systems and deformation scheme of Burtzev-Mikhailov-Zakharov}

The possibility to construct the  class of ``deformed'' integrable
systems,
or integrable systems with ``variable spectral parameter'', 
different from the Ernst equation, was first discovered by   
Burtzev, Mikhailov and Zakharov \cite{BuMiZa}. They
proposed to consider the Lax pairs of the form
\begin{equation}
\frac{\p\Psi}{\p x}=U\Psi\;,\hskip1.0cm
\frac{\p\Psi}{\p y}=V\Psi\;,
\label{lsBMZ}
\end{equation}
where $x$ and $y$ are independent variables; matrices $U$ and $V$ depend on
$(x,y)$ and the ``variable spectral parameter'' $\nu$ (which in turn  depends on
$(x,y)$ and the ``hidden'' spectral parameter ${\l}$):
\baa
U(x,y,\nu)= u_0(x,y)+\sum_{n=1}^{N_1}\frac{u_n(x,y)}{\nu(x,y)-\g_n(x,y)}\;,\\
V(x,y,\nu)= v_0(x,y)+\sum_{n=1}^{N_2}\frac{v_n(x,y)}{\nu(x,y)-\gt_n(x,y)}\;.
\label{UVBMZ}
\eaa
As a part of compatibility conditions of the linear system (\ref{lsBMZ}), after an appropriate fractional-linear transformation in
$\nu$-plane,  the following 
system of equations for $\nu(x,y,{\l})$ must be satisfied:
\begin{equation}
\frac{\p \nu}{\p y}+\sum_{m=1}^{N_2}\frac{b_m}{\nu-\gt_m}=0\;,\hskip1.0cm
\frac{\p\nu}{\p x}+\sum_{m=1}^{N_1}\frac{c_m}{\nu-\g_m}=0\;,
\label{gxyBMZ}
\end{equation}
where $b_n$ and $c_n$ are certain functions of $(x,y)$.
The compatibility condition of the system (\ref{gxyBMZ})  gives the following system for  $\g_n(x,y)$ and $\gt_n(x,y)$:
\begin{equation}
\frac{\p\g_n}{\p y}+\sum_{m=1}^{N_2}\frac{b_m}{\g_n-\gt_m}=0\;,\hskip1.0cm
\frac{\p\gt_n}{\p x}+\sum_{m=1}^{N_1}\frac{c_m}{\gt_n-\g_m}=0\;,
\la{gnBMZ}
\end{equation}
\begin{equation}
\frac{\p c_n}{\p y}-2c_n\sum_{m=1}^{N_2}\frac{b_m}{(\g_n-\gt_m)^2}=0\;,\hskip1.0cm
\frac{\p b_n}{\p x}-2 b_n\sum_{m=1}^{N_1}\frac{c_m}{(\gt_n-\g_m)^2}=0\;.
\la{bcBMZ}
\end{equation}

It is easy to establish the relationship between solutions of the  system
(\ref{gxyBMZ}), (\ref{gnBMZ}), (\ref{bcBMZ}), 
and solutions of  our system (\ref{glln}), (\ref{glmn}), (\ref{almn}).

Namely, suppose that function $\nu({\l},\{{\l}_m\}_{m=1}^M)$
satisfies equations (\ref{glln}) with respect to variables
${\l}_m$. Assume that $M=N_1+N_2$ and split the set of variables
$\{{\l}_1,\dots,{\l}_{N_1+N_2}\}$ into two subsets:  $\{{\l}_1,\dots,{\l}_{N_1}\}$ and
$\{\lt_1,\dots,\lt_{N_2}\}$ where $\lt_n\equiv {\l}_{N_1+n},\; n=1,\dots,N_2$.
In the same way we split the set $\{\g_m\}$ of values of function $\nu(P)$ at these points:  
$$
\{\g_1,\dots,\g_M\}=\{\g_1,\dots,\g_{N_1}\}\cup
\{\gt_1,\dots,\gt_{N_2}\}\;,
$$
where $\gt_n\equiv \g_{N_1+n}$, $n=1,\dots,N_2$.

Now assume that the ``untilded''  variables ${\l}_1,\dots,{\l}_{N_1}$ are arbitrary
functions of variable $x$ and the ``tilded'' variables
$\lt_1,\dots,\lt_{N_2}$
are arbitrary functions of variable $y$.
Then using  (\ref{glln}) we get the derivative of $\nu(P)$ with respect to $x$:
\begin{equation}
\f{\p\nu}{\p x}=\sum_{m=1}^{N_1}\f{\p\nu}{\p{\l}_m}\f{\p{\l}_m}{\p x}=
-\sum_{m=1}^{N_1}\f{\p{\l}_m}{\p x}\f{\a_m}{\nu-\g_m}\;;
\end{equation}
therefore,
\begin{equation}
\frac{\p\nu}{\p x}+\sum_{m=1}^{N_1}\frac{c_m}{\nu-\g_m}=0\;,
\end{equation}
where $c_m\equiv\a_m\f{\p{\l}_m}{\p x}$; this coincides with the second
equation in (\ref{gxyBMZ}). The first equation in (\ref{gxyBMZ}) is
obtained in the same way after identification 
$$b_m\equiv
\a_{N_1+m}\frac{\p{\l}_{N_1+m}}{\p y}\;.$$

Equations (\ref{gnBMZ}) and (\ref{bcBMZ}) for $\g_n$, $b_n$ and $c_n$
as functions of $(x,y)$ arise as compatibility conditions of equations
for $\nu_x$ and $\nu_y$.

Therefore, spaces of rational maps of given degree provide 
solutions of the  system (\ref{gxyBMZ}),
(\ref{gnBMZ}), (\ref{bcBMZ}) if 
\begin{enumerate}\rm
\item
We split the set of the branch points
$\{{\l}_m\}$
into two subsets and 
\item
Assume that one subset contains the branch points which are
arbitrary  functions of $x$ only  and another subset contains the
branch points which are
arbitrary functions of $y$ only.
\end{enumerate}

Concluding, we see that the system (\ref{hie1})  provides a 
 realizations of the deformation scheme of \cite{BuMiZa}.

\subsection{Solutions via the  Riemann-Hilbert problem}

Solutions of  
systems (\ref{hie1}) 
in terms of  matrix
Riemann-Hilbert problem
are given by the following theorem.

\begin{theorem}\la{RH}
Consider  a closed contour $\G$ on the branch covering ${\L}$
 such that its projection
$\pi(\G)$ on the ${\l}$-plane is independent of $\{{\l}_m\}$. Assume that
none of the ramification points $P_m$ belongs to $\G$.
Define a non-degenerate matrix function $\G\ni P\mapsto H(P)$ which is
independent of $\{{\l}_m\}$.
Suppose that a function
$\Psi(P)$ satisfies the following Riemann-Hilbert problem on ${\L}$:
\begin{enumerate}
\item
$\Psi(P)$ is holomorphic and non-degenerate on ${\L}$ outside of the contour
$\G$ where it has the finite  boundary values  related as follows:
\begin{equation}
\Psi_+ (P) = \Psi_- (P) H(P)\;.
\la{RimHil}
\end{equation}
\item
$\Psi$ satisfies the normalization condition at $\infty^{(1)}$ (this
is the  point of ${\L}$ such that   $\pi(\infty^{(1)})=\infty$ and $\nu(\infty^{(1)})=\infty$):
\begin{equation}
\Psi(\infty^{(1)})= I\;.
\la{normir}
\end{equation}
\end{enumerate}
Then the function $\Psi$ satisfies the linear system (\ref{ls}) with
\begin{equation}
G=\Psi(P_0)\;,
\la{GPsi}
\end{equation}
and, therefore, the function $G$ solves the  system (\ref{hie1}).
\end{theorem}

{\it Proof.}
Let $\Psi$ satisfy the Riemann-Hilbert problem. Consider its
logarithmic derivative $\Psi_{{\l}_m}\Psi^{-1}$.
 Due to the independence of the
contour $\G$ and the matrix $H(P)$ of ${\l}_m$, this logarithmic derivative
is single-valued on ${\L}$.  Moreover, it is obviously holomorphic on
${\L}$ outside of the point $P_m$. Let us  write the first two terms of
the Taylor expansion of $\Psi(P)$ near $P_m$:
\ben
\Psi =\Psi_0 +\sqrt{{\l}-{\l}_m}\Psi_1+\dots\;;
\een
then
\ben
\Psi_{{\l}_m}\Psi^{-1} = \f{-1}{2\sqrt{{\l}-{\l}_m}}\Psi_1\Psi_0^{-1} + O(1)\;.
\een
Therefore, in terms of the uniformization map $\nu(P)$, we can write
\ben
\Psi_{{\l}_m}\Psi^{-1}=\f{C_m}{\nu(P)-\g_m}
\een
with some matrix $C_m$ which is independent of $P$. The constant term in this formula is absent due to
normalization condition (\ref{normir}). To compute $C_m$ we put $P=P_0$,
i.e. $\nu(P)=\g_0$; then, using (\ref{GPsi}), we get
\ben
C_m= (\g_0-\g_m)G_{{\l}_m} G^{-1}\;,
\een
which shows that the function $\Psi$, indeed, satisfies (\ref{ls}), and
the corresponding function $G$ solves the system (\ref{hie1}).
$\Box$
\vskip0.5cm

Below we also establish a relationship between systems (\ref{hie1}) and
another type of the Riemann-Hilbert problems, where the function $\Psi$ is
allowed to have regular singularities.

\subsection{Relationship to  isomonodromic deformations}
\la{Reliso}

The set of solutions of each system  (\ref{hie1}) has a subset of solutions 
corresponding to  isomonodromic deformations of  ordinary
differential equations with meromorphic matrix coefficients:
\begin{equation}
\f{d\Psi}{d\g}=A(\g)\Psi\;,\hskip0.8cm A(\g)=\sum_{j=1}^L\f{A_j}{\g-z_j}\;,
\la{Phig}
\end{equation}
where $A_j\in gl(r)$ are certain matrices which are independent of
$\g$ and such that $\sum_{j=1}^L A_j=0$.  Consider $GL(r)$-valued
solution  $\Psi(\g)$ of (\ref{Phig})
satisfying the initial condition at some point $\g_0\in\CP1$:
\begin{equation}
\Psi(\g_0)=I\;.
\la{norm}
\end{equation}
The solution $\Psi(\g)$ has  regular singularities at the points $z_j$; this function is generically non-single-valued
in the $\g$-plane: it gains the right multipliers $M_j$ under
analytical continuation along  
contours starting at $\g_0$ and
 encircling poles $z_j$. The matrices $M_j$ are called the monodromy
matrices of equation (\ref{Phig}); they generate the monodromy group
of equation (\ref{Phig}). If all the monodromy matrices are
independent of the positions of singularities $\{z_j\}$, then (in the
generic case, when  none of the  eigenvalues of each matrix $A_j$
differ by an integer number) the function $\Psi$ satisfies the
following equations with respect to the positions of singularities $z_j$:
\begin{equation}
\f{d\Psi}{d z_j}=\left(\f{A_j}{\g_0-z_j}-\f{A_j}{\g-z_j}\right)\Psi\;.
\la{Phizj}
\end{equation}
Compatibility of equations (\ref{Phig}) and (\ref{Phizj}) is equivalent to the Schlesinger system for the functions $A_j(\{z_k\})$:
\ben
\f{\p A_j}{\p z_k}= \f{[A_j,\,A_k]}{z_j-z_k} - \f{[A_j,\,A_k]}{\g_0-z_k}\;,\hskip0.6cm
j\neq k\; ;
\een
\begin{equation}
\f{\p A_j}{\p z_j}= -\sum_{k\neq j}\left(\f{[A_k,\,A_j]}{z_k-z_j} -
\f{[A_k,\,A_j]}{z_k-\g_0}\right)\;.
\la{Schl}
\end{equation}
The tau-function $\tauh_{JM}$ of the Schlesinger system,  introduced by
Jimbo, Miwa and their collaborators 
\cite{MiwJim78},  is defined by the following system:
\begin{equation}
\f{\p}{\p z_j}\log\tauh_{JM} =  \f{1}{2}{\rm res}|_{\g= z_j}\tr\left(\Psi_\g\Psi^{-1}\right)^2\;;
\hskip0.8cm
\f{\p\tauh_{JM}}{\p\bar{z}_j} = 0\;.
\la{taudef}
\end{equation}

Each solution of the  Schlesinger system induces a solution of  hierarchy (\ref{hie1})  according to the
following theorem:
\begin{theorem}\la{hierSchl}
Consider a solution $\{A_j(\{z_k\})\}$ of the Schlesinger system
 (\ref{Schl}), together with the tau-function $\tauh_{JM}$
 and the corresponding solution $\Psi(\g,\{z_j\})$ of
the linear system (\ref{Phig}), (\ref{Phizj}) normalized by
 (\ref{norm}). Let ${\L}$ be a genus $0$  covering of the ${\l}$-plane
with simple  ramification points $P_1,\dots,P_M$; choose the map
 $\nu(P)$ to satisfy (\ref{ginfty}). 
Consider an arbitrary set of points $P_0$ and $Q_1,\dots,Q_L$ on ${\L}$ such that their projections
${\l}_0$ and
$\mu_1,\dots,\mu_L$ on $\CP1$ (respectively) are independent of the branch points ${\l}_1,\dots,{\l}_M$.
Let us assume that the arguments of the solution and the tau-function of the Schlesinger system  are given by the formulas
\begin{equation}
z_j=\nu(Q_j)\;,\hskip0.7cm \g_0=\nu(P_0)\;.
\la{zgam}
\end{equation}
Then the function
\begin{equation}
\Psi(P,\{{\l}_m\})= \Psi\left(\g,\g_0,\{ z_j\}\right)\Big|_{\g=\nu(P),\,\g_0=\nu(P_0),\,z_j=\nu(Q_j)}
\la{phipsi}
\end{equation}
satisfies the linear system (\ref{ls}) of the hierarchy (\ref{hie1})
with the functions $J_m$ defined by
\begin{equation}
J_m(\{{\l}_n\})=\f{\a_m}{\g_m-\g_0}A(\g_m)\;.
\la{Jmdef}
\end{equation}
Therefore, according to (\ref{JmG}), we have $J_m=G_{{\l}_m} G^{-1}$
for some function $G(\{{\l}_m\})$, which satisfies  the system  (\ref{hie1}).

The corresponding tau-function $\tauh$ is related to the Jimbo-Miwa tau-function as follows:
\begin{equation}
\tauh(\{{\l}_m\})= \prod_{j=1}^L \left(\f{\p\nu}{\p{\l}}(Q_j)\right)^{\tr A_j^2/2}\tauh_{JM}(\{z_j\})
\Big|_{z_j=\nu(Q_j)}\;,
\la{tautauh}
\end{equation}
where the derivative ${\p\nu}/{\p{\l}}$ is given by equation (\ref{glln}).
\end{theorem}

{\it Proof.} Taking the derivative of the function $\Psi(P)$  with
respect to ${\l}_m$ using the chain rule, we get
\ben
\f{\p\Psi(P)}{\p{\l}_m}=\f{\p\Psi}{\p\nu(P)}\f{\p\nu(P)}{\p{\l}_m}+\sum_{j=1}^L\f{\p\Psi}{\p z_j}\f{\p z_j}{\p{\l}_m}=
\sum_{j=1}^L\f{A_j}{\nu(P)- z_j}\f{\a_m}{\g_m-\nu(P)}\Psi - \sum_{j=1}^L\f{A_j}{\nu(P)- z_j}\f{\a_m}{\g_m-z_j}\Psi
\een
\ben
=-\f{\a_m}{\nu(P)-\g_m}\sum_{j=1}^L\f{A_j}{z_j-\g_m}\Psi= \f{\g_0-\g_m}{\nu(P)-\g_m}J_m\Psi\;,
\een
where functions $J_m$ are defined by  (\ref{Jmdef}).

Let us show how to prove the relation  (\ref{tautauh}) between tau-functions.
Taking into account the definition of Jimbo-Miwa tau-functions  (\ref{taudef}),
we have:
\ben
\f{1}{2}\tr\left(\Psi_\g\Psi^{-1}\right)^2=\sum_{j=1}^L\f{\tr A_j^2}{2(\g-z_j)^2} +
\sum_{j=1}^L\f{\p_{z_j}\log\tauh_{JM}}{\g- z_j}\;.
\een
Now, using  definition (\ref{tauhat}) of the tau-function $\tauh$, we get
\begin{equation}
\f{\p\log\tauh}{\p{\l}_m}=\f{\a_m}{2}\tr\left(\Psi_\nu\Psi^{-1}\right)^2 \big|_{P=P_m}=
\sum_{j=1}^L\f{\tr A_j^2}{2}\f{\a_m}{(\g_m-z_j)^2}+\sum_{j=1}^L \f{\a_m}{\g_m-z_j}\p_{z_j}\log\tauh_{JM}\;.
\la{hamham}
\end{equation}
By (\ref{glln}), we see that
\ben
\f{\p}{\p{\l}_m}\log\left\{\f{\p \nu}{\p{\l}}(Q_j)\right\}=\f{\a_m}{(\g_m- z_j)^2}\,,\hskip0.8cm
\f{\p z_j}{\p{\l}_m}= \f{\a_m}{\g_m- z_j}\;;
\een
therefore, applying the chain rule in (\ref{hamham}) (and taking into
account that $\tr A_j^2$ are integrals of the Schlesinger system), we come to (\ref{tautauh}).
$\Box$
\vskip0.5cm
\begin{remark}\rm
The relationship between the systems (\ref{hie1}) and isomonodromic deformations is a generalization of the link between
the Ernst equation  and the Schlesinger system established in  \cite{KorNic96}.
\end{remark}

\subsection{Rank 1  systems}

When the function $G$ from
(\ref{hie1}) is a scalar one,  the  system (\ref{hie1}) can be
rewritten as a system of {\it linear} scalar second order differential equations in terms
of the function
$\fu(\{{\l}_m\})= \log G$:
\begin{equation}
(\g_m-\g_n)^2\f{\p^2\fu}{\p{\l}_m\p{\l}_n}-\a_n\f{\g_m-\g_0}{\g_n-\g_0}\f{\p\fu}{\p{\l}_m}-
\a_m\f{\g_n-\g_0}{\g_m-\g_0}\f{\p\fu}{\p{\l}_n}=0\;,\hskip0.8cm m\neq n\;.
\la{scalar}
\end{equation}
In derivation  of the  system (\ref{scalar}) from (\ref{hie1}) we used 
equations (\ref{glmn}).

In particular, any solution of the matrix system (\ref{hie1}) gives a
solution of the scalar system (\ref{scalar}) if we put $f=\log\det G$.

The linear system (\ref{ls})  turns in rank $1$ into the scalar system
\begin{equation}
\f{\p\psi(P)}{\p{\l}_m}=\f{\g_0-\g_m}{\nu(P)-\g_m}\f{\p\fu}{\p{\l}_m}\;,
\la{lsscal}
\end{equation}
where $\psi(P,\{{\l}_m\})=\log\Psi$. As well as in the 
matrix case, the function $\psi$ is generically non-single-valued on ${\L}$.

The definition of tau-function (\ref{tauhat}) now looks as
follows:
\begin{equation}
\f{\p\log\tauh}{\p{\l}_m}= \f{(\g_0-\g_m)^2}{2\a_m}\left\{\f{\p\fu}{\p{\l}_m}\right\}^2\;.
\end{equation}
Alternatively, it can be rewritten in terms of the function
$\psi(P)$, using (\ref{tauhat2}):
\begin{equation}
\f{\p\log\tauh}{\p{\l}_m}=\f{1}{2}\res\big|_{P_m}\left\{\f{(d\psi)^2}{d{\l}}\right\} \;,
\la{tauhat3}
\end{equation}
where $d\psi(P)=\psi_\nu(P) d\nu(P)$.

Let us discuss the solutions of system (\ref{lsscal}) and equation
(\ref{scalar}).
\begin{theorem}
Let $\G$ be an arbitrary  smooth closed contour on ${\L}$ such that its
projection on the ${\l}$-plane $\pi(\G)$ is independent of $\{{\l}_m\}$ and
$P_m\not\in \G$ for any $m$.
Consider on $\G$ an arbitrary H{\"o}lder-continuous function $h(P)$ independent
of $\{{\l}_m\}$. Then the function
\begin{equation}
f =\oint_{\G}\f{h(Q)d\nu(Q)}{\nu(Q)-\g_0}
\la{solscal}
\end{equation}
satisfies system (\ref{scalar}).
\end{theorem}
{\it Proof.}
In the scalar  case we know explicitly the solution of an arbitrary  Riemann-Hilbert
problem from
theorem \ref{RH}.  If we define the  function
\ben
 h(P)=\f{1}{2\pi i}\log H(P)\;
\een
(we assume that it is single-valued on $\G$ i.e. the index of $H(P)$ on
$\G$ equals $0$),
the Riemann-Hilbert problem (\ref{RimHil}) becomes additive:
\begin{equation}
\psi_+ (P) = \psi_- (P) + 2\pi i h(P)
\la{scalRH}
\end{equation}
for $P\in\G$
(we recall that $\psi=\log\Psi$); here  $\psi_\pm$ stands for the
boundary values of the
function $\psi$ on $\G$.  Normalization condition (\ref{normir})
turns into
\begin{equation}
\psi(\infty^{(1)})=0\;.
\la{normir1}
\end{equation}
The solution of (\ref{scalRH}), (\ref{normir1}) for
a H{\"o}lder-continuous functions $h(P)$ is given by the Cauchy integral 
\begin{equation}
\psi(P)=\oint_{\G}\f{h(Q)d\nu(Q)}{\nu(Q)-\nu(P)}\;,
\la{solscal1}
\end{equation}
which implies (\ref{solscal}) at $P=P_0$, when $\nu(P)=\g_0$.
%Changing the variable of integration to ${\l}$, we also get
%\begin{equation}
%\psi(P)=\oint_{\G}\f{h(\g)}{\g({\l})-\g(P)}\f{d\g}{d{\l}}d{\l}\;.
%\la{solscal1}
%\end{equation}
$\Box$
\vskip0.5cm

 Formula (\ref{solscal}) can also be verified  as follows. First,
one can check by the direct substitution, using equations (\ref{glln}),
(\ref{glmn}), (\ref{almn}), that the function
\begin{equation}
f  =\f{1}{\nu(Q)-\g_0}\f{\p\nu(Q)}{\p{\l}}
\la{solelem}
\end{equation}
satisfies  system (\ref{scalar}) for any ${\l}\equiv \pi(Q)$ independent of $\{{\l}_m\}$.
Using linearity of equation (\ref{scalar})   we can consider
the superposition of these solutions at different $Q\in\G$ with an arbitrary
$\{{\l}_m\}$-independent measure $h(Q)$, which leads to (\ref{solscal}).

In fact, we can consider any (say, compact) subset $D\subset{\L}$ whose
projection on ${\l}$-plane is $\{{\l}_m\}$-independent, and such that
$P_m\not\in D$.
We can define an
arbitrary $\{{\l}_m\}$-independent measure $d\mu (Q)$ on $D$; then
the superposition
principle implies that the function
\begin{equation}
f =\int_{D}\f{h(Q)}{\nu(Q)-\g_0}\f{\p\nu(Q)}{\p{\l}} d\mu(Q)
\la{solgen}
\end{equation}
is a solution of (\ref{scalar}).

\vskip0.3cm

{\bf The Euler-Darboux equation.}
As we noticed earlier,
for the twofold covering with two branch points
${\l}_1=\xi$ and ${\l}_2=\xibar$ the system (\ref{hie1}) is
equivalent to Ernst equation (\ref{Ernst}). In scalar case  we get
 the following equation in terms of
$f =\log G$:
\ben
f_{{\xi\xibar}}-\f{f_{\xi}-f_{\xibar}}{2(\xi-\xibar)}=0\;.
\een
If we introduce the real coordinates $(z,\rho)$ such that $\xi=z+ i\rho$,
$\xibar= z-i\rho$, this equation takes the form of
Euler-Darboux equation:
\begin{equation}
f_{zz}+\f{1}{\rho}{f_\rho}+f_{\rho\rho}=0\;.
\la{ED}
\end{equation}
From (\ref{unifor}), taking into account that $P_0=\infty^{(2)}$ and
$\g_0=({\l}_1+{\l}_2)/2$, we obtain:
\ben
\f{1}{\nu(P)-\g_0}\f{\p\nu(P)}{\p{\l}} = \f{1}{[({\l}-\xi)({\l}-\xibar)]^{1/2}}\;.
\een
Then representation  (\ref{solscal}) gives the solution
\cite{HilKur} of the Euler-Darboux equation:
\ben
f =\oint_\G \f{h({\l})d{\l}}{[({\l}-\xi)({\l}-\xibar)]^{1/2}}\;.
\een

\section{Generalization to  higher genus}

Here we discuss possible ways to define systems, analogous to
(\ref{hie1}), starting from 
 Hurwitz spaces $H_{g,N}$ (which are spaces of meromorphic functions
of degree $N$ over Riemann surfaces of genus $g$) in  genus $g\geq 2$ (we skip the  case $g=1$
when the classification of stable bundles is rather special \cite{Atiyah}, see also remark \ref{gen1} below).

As before, denote the critical points
of
a meromorphic function $\pi(P)$ on a Riemann surface of genus $g$ by $P_1,\dots,P_M$ and their images in $\CP1$
by ${\l}_m=\pi(P_m)$. This meromorphic function realizes the Riemann
surface as a   ramified $N$-fold covering ${\L}$
of genus $g$ of the ${\l}$-sphere  with ramification points $P_1,\dots,P_M$.
Assuming that all the branch points
${\l}_m$ are different and  simple, we get,
according to the Riemann-Hurwitz formula:
\ben
M=2g+2N-2\;.
\een
Let's introduce on ${\L}$ a canonical  basis of cycles $(a_\a, b_\a)$
($\a=1,\dots,g$) and corresponding  basis of holomorphic 1-forms
$w_\a(P)$ ($\a=1,\dots,g$), normalized by $\oint_{a_\a}w_{\b}=\delta_{\a\b}$.

\subsection{Rank 1  systems in arbitrary genus}

The scalar systems (\ref{scalar}) admit natural generalization to the
Hurwitz spaces $H_{g,N}$.
Denote the prime-form on ${\L}$ by $E(P,Q)$ (where $P,Q\in{\L}$); introduce the Bergmann
kernel $B(P,Q)=d_Pd_Q\log E(P,Q)$.
 By $B_m(P)$ we denote the meromorphic
differential of 2nd kind with vanishing $a$-periods and single pole at $P_m$
of the second order with the following local  behaviour:
\ben
B_m(P)=\left(\f{1}{\lp_m^2}+O(1)\right)d\lp_m
\een
as $P\to P_m$,
where $\lp_m=\sqrt{{\l}-{\l}_m}$ is a local parameter near
$P_m$. Differentials $B_m$ are related to the Bergmann kernel as follows:
\begin{equation}
B_m(P)=\f{B(P,Q)}{d\lp_m(Q)}\Big|_{Q=P_m}\;.
\la{Wmberg}
\end{equation}
Corresponding abelian integrals we denote by ${\O}_m$:
\begin{equation}
{\O}_m(P)=\int_{Q_0}^P B_m \equiv \f{d}{d\lp_m (Q)}\left\{\log\f{E(P,Q)}{E(Q_0,Q)}\right\}\Big|_{Q=P_m}\;,
\la{OmP}
\end{equation}
where $Q_0\in {\L}$ is a base-point such that its projection  on
${\l}$-plane does not depend on $\{{\l}_n\}$.

Let us prove the following variational formula.

\begin{theorem}\la{glm}
Assume that the
local parameters $\lp_P$, $\lp_Q$ don't depend on (some)
${\l}_m$. Introduce the symmetric function
\begin{equation}
b(P,Q)=\f{B(P,Q)}{d\lp_P d\lp_Q}\equiv \f{\p^2}{\p\lp_P\p\lp_Q}\log E(P,Q)\;.
\la{gPQ}
\end{equation}
Then dependence of  $b(P,Q)$ on the branch point ${\l}_m$ is given by
the following equation:
\begin{equation}
\f{\p b(P,Q)}{\p{\l}_m} =\f{1}{2} b(P,P_m) b(Q,P_m)
\la{derglm}
\end{equation}
\end{theorem}
{\it Proof.}
Formula (\ref{derglm}) is closely related to the Rauch variational
formulas \cite{Rauch,Fay92} which describe dependence of holomorphic
differentials on the branch points. The proof is also very similar.
Namely, consider the ${\l}_m$-derivative of the Bergmann kernel:
${\p B(P,Q)}/{\p{\l}_m}$. This is a symmetric 1-form on ${\L}\times
{\L}$.
Consider the Taylor series of $B(P,Q)$ with respect to its first
argument $P$  in a neighbourhood of ramification point $P_m$:
\begin{equation}
B(P,Q)= \{a_0+ a_1\lp_m +...\}d\lp_m=\{a_0+ a_1\lp_m +...\}\f{d{\l}}{2\sqrt{{\l}-{\l}_m}}\;,
\la{Sloc}
\end{equation}
where $a_0, a_1,\dots$ are some 1-forms with respect to $Q$.
Differentiation of (\ref{Sloc}) with respect to ${\l}_m$ gives
\begin{equation}
\f{ \p B(P,Q)}{\p{\l}_m}= \{a_0+ o(1)\}\f{d{\l}}{4({\l}-{\l}_m)^{3/2}}=\{a_0+ o(1)\}\f{d\lp_m}{2\lp_m^2}\;.
\la{Wloc}
\end{equation}
as $P\to P_m$.
 Therefore,  ${\p B(P,Q)}/{\p{\l}_m}$  is a  meromorphic 1-form with
respect to $P$, with the only pole at $P_m$
of the second order with leading coefficient 
\ben
\f{a_0(Q)}{2}\equiv\f{B(Q,P)}{2 d\lp_m(P)}\Big|_{P=P_m}
\een
 (which is itself a
1-form with respect to $Q$) and vanishing $a$-periods. 
Taking into account the  symmetry of $B(P,Q)$, we get
\begin{equation}
\f{ \p B(P,Q)}{\p{\l}_m}=\f{1}{2}\left\{\f{B(P,R)}{d\lp_m(R)}\Big|_{R=P_m}\right\}
\left\{\f{B(Q,R)}{d\lp_m(R)}\Big|_{R=P_m}\right\}
\end{equation}
which is equivalent to (\ref{derglm}).
$\Box$

Since differentials $B_m$ are related to the Bergmann kernel via
(\ref{Wmberg}), 
the  theorem immediately implies the following useful
\begin{corollary}\la{WOlm}
The abelian differential $B_n(P)$ and the abelian  integral ${\O}_n(P)$ depend
on ${\l}_m$ (for any $m\neq n$) as follows: 
%(assuming that projection of
%their  argument  $P$ on ${\l}$-plane
%is ${\l}_m$-independent):
 \begin{equation}
\f{\p}{\p{\l}_m} B_n(P) = \f{1}{2} b(P_m,P_n) B_m(P)\;,\hskip1.0cm 
\f{\p}{\p{\l}_m} {\O}_n(P) = \f{1}{2} b(P_m,P_n) {\O}_m(P)\;.
\la{OWlm}
\end{equation}
\end{corollary}

\vskip0.3cm
Now we are in position to formulate the  analogs of Euler-Darboux
equation in arbitrary genus:

\begin{theorem}\la{genus}  
Consider the following linear system of scalar equations for function
$\psi(P,\{{\l}_m\})$:
\begin{equation}
\f{d\psi(P)}{d{\l}_m}= R_m {\O}_m(P)\;,
\la{lsgenus}
\end{equation}
where $R_m$ are some functions of $\{{\l}_m\}$. Denote by $P_0$ a
point of ${\L}$ such that ${\l}_0\equiv \pi (P_0)$ does not depend on 
$\{{\l}_m\}$. 
Then the compatibility conditions of the system (\ref{lsgenus}) are given
by the following equations: 
\begin{equation}
\f{\p^2 f}{\p{\l}_m\p{\l}_n}-\f{b(P_m,P_n)}{2}
\left\{\f{v_n}{v_m}\f{\p f}{\p{\l}_m}+
\f{v_m}{v_n}\f{\p f}{\p{\l}_n}\right\}=0\;,
\la{EDgenus}
\end{equation}
where $f(\{{\l}_m\})=\psi(P_0)$ and
\ben
v_m\equiv {\O}_m(P_0)=\int_{Q_0}^{P_0} B_m\;;
\een
function $b(P,Q)$ is defined by (\ref{gPQ}).
\end{theorem}
{\it Proof.}
Substituting in (\ref{lsgenus}) $P=P_0$, we see that 
\ben
R_m= \f{f_{{\l}_m}}{{\O}_m(P_0)}\;.
\een
Then the compatibility conditions of the system (\ref{lsgenus}) are
equivalent to the system of equations 
\begin{equation}
 \f{\p}{\p{\l}_m}\left\{\f{\p f}{\p{\l}_n}\f{{\O}_n(P)}{{\O}_n (P_0)}\right\}-
\f{\p}{\p{\l}_n}\left\{\f{\p f}{\p{\l}_m}\f{{\O}_m(P)}{{\O}_m
(P_0)}\right\}=0\;,\hskip0.8cm m\neq n
\la{compgenus}
\end{equation}
Using (\ref{OWlm}), we rewrite this equation as follows:
\ben
f_{{\l}_m{\l}_n}\left\{\f{{\O}_n(P)}{{\O}_n(P_0)}-\f{{\O}_m(P)}{{\O}_m(P_0)}\right\}
\een
\begin{equation}
+\f{b(P_m,P_n)}{2}\left\{{\O}_m(P){\O}_n(P_0)-{\O}_n(P){\O}_m(P_0)\right\}\left\{\f{f_{{\l}_n}}{{\O}_n^2(P_0)}+
\f{f_{{\l}_m}}{{\O}_m^2(P_0)}\right\}=0\;.
\la{compgen1}
\end{equation}
The  l.h.s. of this equation is an abelian integral on ${\L}$
with respect to argument $P$ with vanishing $a$-periods (since all
$a$-periods of differentials $B_m(P)$ vanish) and the poles (of the first order)
only at $P_m$ and $P_n$.
Since ${\O}_m(P)= -1/\lp_m+O(1)$ as $P\to P_m$, we
immediately see that equations (\ref{EDgenus}) are equivalent to
absence of poles of  (\ref{compgen1}) at $P_m$ and $P_n$.
Therefore, the l.h.s. of (\ref{compgen1}) must be a constant with
respect to $P$; choosing $P=P_0$ we see that this constant vanishes.
We conclude that equations (\ref{EDgenus})  indeed provide 
compatibility of the linear system (\ref{lsgenus}).

$\Box$

In analogy to the case of genus zero, now we shall define the
tau-function of the system (\ref{EDgenus}) and construct its solutions
via solutions of the scalar Riemann-Hilbert problem on ${\L}$.

We   define the tau-function of the  system
(\ref{EDgenus}) by the following system of equations:
\begin{equation}
\f{\p}{\p{\l}_m}\log\tauh=\f{f_{{\l}_m}^2}{v_m^2}\;,\hskip0.7cm 
\f{\p\log\tauh}{\p\overline{{\l}_m}}=0\;.
\la{deftaugen}
\end{equation}
From variational formulas (\ref{OWlm}) and equations (\ref{EDgenus})
it follows that 
\begin{equation}
\f{\p}{\p{\l}_n}\left\{\f{f_{{\l}_m}^2}{v_m^2}\right\}= b(P_m,P_n)\f{f_{{\l}_m}f_{{\l}_n}}{v_m
v_n}\;.
\la{tausecond}
\end{equation}
Symmetry of this expression with respect to $m$ and $n$ proves
compatibility of equations (\ref{deftaugen}).

As well as in the case of genus zero,  we can prove
the following
\begin{lemma}
Definition (\ref{deftaugen}) of the tau-function can be alternatively
rewritten in the following form:
\begin{equation}
\f{\p\log\tauh}{\p{\l}_m}=\f{1}{2}\res\Big|_{P_m}\left\{\f{(d_P\psi)^2}{d{\l}}\right\}\;.
\la{def1genus}
\end{equation}
\end{lemma}
{\it Proof.} Choosing the standard local parameter
$\lp_m=\sqrt{{\l}-{\l}_m}$ near $P_m$, we have $d\lp_m=d{\l}/2\lp_m$;
therefore,
\ben
\res\Big|_{P_m}\left\{\f{(d_P\psi)^2}{d{\l}}\right\}=\f{1}{2}\left(\f{\p\psi}{\p\lp_m}\right)^2(P_m)\;.
\een
On the other hand, as $P$ belongs to a neighbourhood of $P_m$, the
linear system (\ref{lsgenus}) can be rewritten as follows:
\ben
\f{\p\psi}{\p{\l}_m}-\f{1}{2\lp_m}\f{\p\psi}{\p\lp_m}=\f{\p f}{\p{\l}_m}\f{{\O}_m(P)}{{\O}_m(P_0)}\;,
\een
where we separated the
dependence of $\psi$ on ${\l}_m$ which comes from ${\l}_m$-dependence of
$\lp_m$.
Taking the residue at $P_m$, we get
\ben
\f{\p\psi}{\p\lp_m}(P_m)=2\f{f_{{\l}_m}}{{\O}_m(P_0)}\;,
\een
which implies  coincidence of (\ref{deftaugen}) and (\ref{def1genus}).

$\Box$

The next theorem provides solutions of the system (\ref{EDgenus}).

\begin{theorem}
Let $\G$ be an arbitrary  smooth closed contour on ${\L}$ such that its
projection on the ${\l}$-plane $\pi(\G)$ is independent of $\{{\l}_m\}$ and
$P_m\not\in \G$ for any $m$. 
Consider on $\G$ an arbitrary H{\"o}lder-continuous  function $h(Q)$ independent
of $\{{\l}_m\}$. 

Then the function
\begin{equation}
f =\oint_{\G} h(Q) d_Q\log\f{E(P_0,Q)}{E(Q_0,Q)}    
\la{solgenus}
\end{equation}
satisfies the  system
(\ref{EDgenus}). Corresponding solution of the linear system
(\ref{lsgenus}) is given by
\begin{equation}
\psi(P) =\oint_{\G} h(Q) d_Q\log\f{E(P,Q)}{E(Q_0,Q)}  \;.      
\la{sollsgenus}
\end{equation}
\end{theorem}
{\it Proof.}
Before proving the statement of the theorem we notice that if
the function $h(Q)$ is H{\"o}lder-continuous, then
$\psi$ (\ref{sollsgenus}) is a solution of the Riemann-Hilbert problem
on the
contour $\G$ with jump $2\pi i h(Q)$. 

Let us  now  verify that the functions $f$ and $\psi$ defined by
(\ref{solgenus}) and (\ref{sollsgenus}), satisfy the linear system
(\ref{lsgenus}):
\begin{equation}
\psi_{{\l}_m}=\f{{\O}_m(P)}{{\O}_m(P_0)}f_{{\l}_m}\;.
\la{lsaux}
\end{equation}
 
Taking into account
representation (\ref{OmP}) of ${\O}_m(P)$ 
 in terms of  prime-forms, in analogy to Rauch formulas
given by Th.\ref{glm}, we see that
the Cauchy kernel  $d_Q\log\f{E(P,Q)}{E(Q_0,Q)}$ depends as follows on ${\l}_m$
(assuming that all points $P,Q,Q_0$ are ${\l}_m$-independent):
\begin{equation}
\f{\p}{\p{\l}_m}\left\{d_Q\log\f{E(P,Q)}{E(Q_0,Q)}\right\}=\f{1}{2}B_m(Q){\O}_m(P)\;.
\la{Cauchylm}
\end{equation}

Therefore,
\ben
\f{\psi_{{\l}_m}(P)}{{\O}_m(P)}=\f{1}{2}\oint_{\G}h(Q) B_m(Q)\;.
\een
Independence of this expression of $P$ proves
(\ref{lsaux}). Therefore, the function $f$ satisfies the compatibility
conditions of (\ref{lsaux}) i.e. the 
equations (\ref{EDgenus}).

$\Box$

We notice that, using superposition principle for the
 systems (\ref{EDgenus}), we can substitute integration
over contour in (\ref{solgenus}) by integration with respect to  an
 arbitrary $\{{\l}_m\}$-independent measure with compact support; 
the only 
condition we need to impose on this measure is commutativity of integration in
(\ref{solgenus})
with differentiation with respect to $\{{\l}_m\}$.

\subsection{Systems of higher rank}

Here we consider coverings of genus $g\geq 2$.
To define a  natural rank $r$ analogs of  equations (\ref{EDgenus}),  
it is natural to start
from generalization of the zero curvature representation (\ref{ls}) to higher genus:
\begin{equation}
\frac{\p\Psi}{\p{\l}_m}= U_m\Psi\;,
\la{lshr}
\end{equation}
where the $r\times r$ ``Lax matrix'' $U_m(P,\{{\l}_n\})$, $P\in{\L}$   has
only one singularity on ${\L}$, 
which is a simple pole at
the ramification point $P_m$. We shall assume that $\tr U_m =0$  (this
condition can be always satisfied by  normalization of function $\Psi$:
$\Psi\to [\det \Psi]^{-1/r}\Psi$).
As well as in rank $1$, the ``Lax matrix'' $U_m(P)$ can not be
single-valued on ${\L}$. In rank $1$  matrix $U_m$ has additive twists
along basic cycles on ${\L}$, but $d_P U_m(P)$ is a single-valued
meromorphic differential on ${\L}$ with pole of second order at $P_m$. 

%In contrast to genus zero case, $U_m$ can not be simply a meromorphic  function (i.e. section of a trivial
%bundle) over ${\L}$; instead, $U_m(P)$ must have some kind of multiplicative and (or) additive non-singlevaluedness under the
%tracing along topologically non-trivial closed contours  on ${\L}$. 
In higher rank this is not enough -  it is necessary to introduce more
degrees of freedom, and allow $d_PU_m$ to suffer similarity
transformation under tracing along each topologically non-trivial closed contour on ${\L}$ (this
transformation may be $\{{\l}_m\}$-dependent itself). 

More precisely,
consider a stable vector bundle $\hi$ of rank $r$ and degree $d$ over
${\L}$. Require $d_PU_m$ to be a meromorphic section of the bundle
$ad_{\hi}\otimes K$ with quadratic pole at the ramification  point $P_m$ ($K$
is the canonical line bundle), i.e. 
$$d_PU_m\in H^0({\L},ad_\hi\otimes K (2P_m))\;;$$
according to terminology proposed by Hitchin \cite{Hitchin}, $d_PU_m$ is
called the meromorphic Higgs field.

According to Narasimhan-Seshadri theorem \cite{NarSes}, the stable
bundle $\hi$ is  characterized by the set of unitary
constant (independent of a  point $P\in {\L}$) matrices
 $\hi_{a_1},\dots,\hi_{a_g},\-
\hi_{b_1},\dots,\hi_{b_g}$, satisfying one relation
$$\prod_{\a=1}^g
\hi_{a_j}\hi_{b_j}\hi_{a_j}^{-1}\hi_{b_j}^{-1}=\exp\{2\pi i
\frac{d}{r}\}\;.$$
Stability of the bundle $\hi$ implies  $h^0(ad_\hi)=0$ i.e.
the bundle $ad_\hi$ does not  have
any holomorphic section. On the other hand,  $h^0(ad_\hi\otimes
K)=(g-1)(r^2-1)$; therefore, fixing the singular part of $d_P U_m$ at
the ramification point $P_m$:
\begin{equation}
d_P U_m(P) = \left(-\frac{R_m}{x_m^2} + O(1)\right)
d x_m\;\hskip0.5cm {\rm as}\hskip0.5cm P\to P_m\;,
\end{equation}
we define $d_P U_m$ up to an arbitrary linear combination of
$(g-1)(r^2-1)$ holomorphic sections of $ad_\hi\otimes K$. In rank 1
case we fixed $d_P U_m$ uniquely  imposing the normalization condition of
vanishing of all $a$-periods of $d_P U_m$; unfortunately, such simple
normalization in higher rank is not known to the authors.

The ``Lax matrix'' $U_m$ is uniquely defined by $d_P U_m$ inside of
the fundamental polygon $\hat{\L}$ of ${\L}$ if we fix a normalization point
$P_0\in\hat{\L}$ and assume that $U_m(P_0)=0$. The only singularity of the
matrix $U_m$ is the simple pole
at $P_m$ of the form
\ben
U_m(P)=\frac{R_m}{x_m}+ O(1)\hskip0.5cm {\rm as}\hskip0.5cm P\to P_m\;;
\een
generically $U_m$ has both multiplicative and additive
non-single-valuedness along any cycle $\gamma\in \pi_1({\L})$:
\begin{equation}
U_m(P^{\gamma})=\hi_{\gamma} U_m(P) \hi_{\gamma}^{-1} + C_m^{\gamma}\;;
\la{UQ}
\end{equation}
since $d_PU_m(P^{\gamma})=
\hi_{\gamma}d_P U_m(P) \hi_{\gamma}^{-1}$, the ``additive twists''
$C_m^{\gamma}$ don't depend on $P$.

The compatibility conditions of the linear system
(\ref{lshr}) are given by:
\begin{equation}
F_{mn}(P)=0\;,
\la{comp}
\end{equation}
where $F_{mn}(P)$ is  the curvature:
\begin{equation}
F_{mn}(P)=\f{\p U_m}{\p{{\l}_n}}-\f{\p U_n}{\p{{\l}_m}}+[U_m\;,\;U_n]\;.
\la{Fmn}
\end{equation}
To rewrite the compatibility conditions (\ref{comp}) in terms of
variables depending only on $\{{\l}_m\}$ (and not on the point $P$ of the
covering ${\L}$), we consider the coefficients of the Taylor series of $U_m(P)$ at the
ramification point $P_n$, $n\neq m$:
\ben
U_m(P)= S_{mn} + T_{mn}x_n + O(x_n^2)\;.
\een
Then non-singularity of  $F_{mn}$ at $P_m$ is
equivalent to the condition
\begin{equation}
\f{\p R_m}{\p{\l}_n} +\f{1}{2}T_{nm} +[R_m,S_{nm}]=0\;,\hskip0.5cm m\neq n\;.
\la{E1}
\end{equation}
The non-singularity of $F_{mn}$ is insufficient for its vanishing,
since $F_{mn}(P)$ has  both multiplicative and
additive twists along topologically non-trivial loops. However, if we require that the  ``additive twists''
$C_m^{\g}$ from (\ref{UQ}) are related to  matrices $\hi_\g$
via equations
\begin{equation}
\f{\p\hi_\g}{\p{\l}_m}\hi_\g^{-1}=C_m^{\g}\;,
\la{E2}
 \end{equation}
we observe that transformation (\ref{UQ}) of the Lax matrices $U_m$ is
 nothing but the gauge transformation of connection 1-form
 $\sum_{m=1}^M U_m d{\l}_m$ by the matrix $\hi_\g(\{{\l}_m\})$. Then the
 curvature coefficient $F_{mn}$  transforms as follows:
\ben
F_{mn}(P^\g)=\hi_\g F_{mn}(P)\hi_\g^{-1}
\een
i.e. $F_{mn}\in H^0(ad_\hi)$; thus $F_{mn}=0$.

We conclude that the compatibility conditions of the linear system
(\ref{lshr})  in arbitrary rank
$r$ and any genus $g\geq 2$ are given by the system of equations
(\ref{E1}) and (\ref{E2}). The new feature in comparison with the 
genus zero case is  that we get another degree of freedom -  the stable bundle $\hi$ must itself depend on 
the branch points ${\l}_m$ according to equations (\ref{E2}) (since
generic vector bundle over ${\L}$ is stable, generically evolution
(\ref{E2}) preserves stability of $\hi$ at least locally, in a
neighbourhood of a given stable bundle).

Obviously, without any normalization of the twisted 1-form $d_PU_m$
the coefficients $S_{mn}$, $T_{mn}$ and additive twists $C_m^{\g}$ are
not uniquely determined by the set of residues $R_m$ and matrices
$\hi_\gamma$. Therefore, the number of equations (\ref{E1}),(\ref{E2})
is substantially smaller than the number of variables.

A possible way to define $d_P U_m$ uniquely is to make use of one of meromorphic
bidifferentials $W(P,Q)$ on ${\L}\times{\L}$ whose existence is
provided by the following lemma:
\begin{lemma}
There exists meromorphic
bidifferential $W(P,Q)$ on ${\L}\times{\L}$ satisfying the following
conditions:
\begin{enumerate}
\item
On the diagonal $P=Q$ the bidifferential $W(P,Q)$ has  second order pole with
biresidue equal to $r^2\times r^2$  matrix $\Pi$ (which is
the permutation  matrix in $\C^r\otimes \C^r$):
\begin{equation}
W(P,Q)=\left\{\frac{\Pi}{(x_P- x_Q)^2} + O(1)\right\}\;
dx_P dx_Q \;.
\la{diag}
\end{equation}
\item
Symmetry condition:
\begin{equation}
W(P,Q)=\Pi W(Q,P)\Pi\;.
\la{symmetry}
\end{equation}

\item
Automorphy conditions: for any $\g\in\pi_1({\L})$ we have
\begin{equation}
W(P^\g,Q)= \stackrel{1}{\hi}_\g W(P,Q)\stackrel{1}{\hi}_\g^{-1}\;,
\la{autom1}
\end{equation}
\begin{equation}
W(P,Q^\g)= \stackrel{2}{\hi}_\g W(P,Q)\stackrel{2}{\hi}_\g^{-1}\;,
\la{autom2}
\end{equation}
where for any linear operator  $A$ in $\C^r$ we denote by
$\stackrel{1}{A}$ and  $\stackrel{2}{A}$ the operators $A\otimes I$
and $I\otimes A$ in $\C^r\otimes \C^r$, respectively.

\end{enumerate}
\end{lemma}
\vskip0.3cm
The relation (\ref{autom2}) is obviously a corollary of
(\ref{autom1}) and the symmetry requirement
(\ref{symmetry}). Equivalently, relations (\ref{autom1}), (\ref{autom2}) mean that $W(P,Q)$ belongs
to $H^0(\stackrel{1}{ad_\hi}\otimes K\,(2Q))$ with respect to its
first argument, and to $H^0(\stackrel{2}{ad_\hi}\otimes K\,(2P))$  with respect to its second argument.

{\it Proof.}
Existence of  bidifferential $W_0(P,Q)$  satisfying only (\ref{diag})
and automorphy properties (\ref{autom1}), (\ref{autom2}),  can be
proved similarly  to, say, existence of Schiffer kernel
corresponding to the 
 bundle $ad_\hi$ 
(see  for example \cite{Fay92}). To construct bidifferential $W(P,Q)$ which satisfies
in addition the symmetry  condition (\ref{symmetry}), suppose that
$F(P,Q)\equiv W_0(P,Q)- \Pi W_0(Q,P) \Pi$ does not vanish; this is a
holomorphic section of $\stackrel{1}{ad_\hi}\otimes K$ with respect to
$P$ and holomorphic section of $\stackrel{2}{ad_\hi}\otimes K$ with
respect to $Q$. Obviously,
$$
\Pi F(P,Q)\Pi = -F(Q,P)\;.
$$
Now define $W(P,Q)= W_0(P,Q)-\frac{1}{2}F(P,Q)$. Simple calculation
shows that it satisfies (\ref{symmetry}); other required properties
are inherited from $W_0(P,Q)$.

$\Box$

We proved existence of $W(P,Q)$;  obviously this bidifferential is not unique: we can add to $W(P,Q)$ an arbitrary  linear
combination of bilinear products of holomorphic sections $f_k(P)$ of
$ad_\hi\otimes K$, satisfying the symmetry condition (\ref{symmetry});
this is a linear combination of the form
$$
\sum_{j,k=1}^{(g-1)(r^2-1)} \alpha_{jk}
\left\{\stackrel{1}{f}_k(P)\stackrel{2}{f}_j(Q)+
\stackrel{1}{f}_j(P)\stackrel{2}{f}_k(Q)\right\}
$$  
with arbitrary $\a_{jk}\in\C$.

Let us fix the bidifferential  $W(P,Q)$ in some way (for example,
according to remark \ref{rema} below). Then the 
Higgs fields $d_P U_m$ can be defined as follows:
\begin{equation}
d_PU_m(P)=\frac{\stackrel{1}{\tr}\left\{\stackrel{12}{W}(P,Q)\stackrel{2}{R}_m\right\}}{dx_m(Q)}\Big|_{Q=P_m}\;.
\la{dPUm}
\end{equation}

Now all variables $S_{mn}$, $T_{mn}$, $C_m^\g$ become functionals of
the residues $R_m$, branch points ${\l}_m$ and  matrices
$\hi_\g$; the number of variables $(\hi_\g,R_m)$ in the system
(\ref{E1}), (\ref{E2}) coincides with the number of equations.

The natural definition of tau-function, in agreement with
(\ref{tauhat2}), looks as follows:
\begin{equation}
\frac{\p}{\p{\l}_m}\log\tau =\frac{1}{2}\res\big|_{P_m}\frac{\tr
(d_P\Psi\Psi^{-1})^2}{d{\l}}\;.
\la{deftauhr}
\end{equation}
Consistency of the definition (\ref{deftauhr}) is provided by the
following
\begin{lemma}
Let function $\Psi$ solve the linear system (\ref{lshr}) with Lax
matrices $U_m$ satisfying condition $U_m(P_0)=0$ for some $P_0\in{\L}$, and conditions (\ref{dPUm}). Then the
1-form 
\begin{equation}
\sum_{m=1}^M \left\{\res\Big|_{P_m}\frac{\tr (d_P\Psi\Psi^{-1})^2}{ d{\l}}\right\}d{\l}_m
\end{equation}
is closed.
\end{lemma}
{\it Proof.} Consider the Taylor series of $\Psi(P)$ in a neighbourhood
of $P_m$:
$$
\Psi(P)= \Psi_0 +x_m \Psi_1 + O(x_m^2)\;;
$$
then
$$
\Psi_{x_m}\Psi^{-1}=\Psi_1\Psi_0^{-1} +O(1)\;,
$$
and
$$
\Psi_{{\l}_m}\Psi^{-1}=-\frac{1}{2x_m}\Psi_1\Psi_0^{-1} +O(1)\;.
$$
Therefore,
$$
R_m=-\frac{1}{2}\Psi_1\Psi_0^{-1}
$$
and
$$
\frac{1}{2}\res\big|_{P_m}\frac{\tr
(d_P\Psi\Psi^{-1})^2}{d{\l}}=\frac{1}{2}\res\big|_{P_m}\frac{\tr(-2 R_m
dx_m)^2}{2x_m dx_m}=\tr R_m^2
$$
Now we have to make sure that the derivative $\frac{\p}{\p{\l}_n}\tr
R_m^2$ is symmetric under the interchanging of  $m$ and $n$:
\begin{equation}
\frac{\p}{\p{\l}_n}\tr R_m^2 = -2\tr
R_m\left\{\frac{1}{2} T_{nm} +[R_m, S_{nm}]\right\} =-\tr R_m T_{nm}\;.
\la{Rmln}
\end{equation}
As $P\to P_m$, we have:
\ben
\tr\left\{d_PU_m(P) d_PU_n(P)\right\}= (-\tr R_m T_{nm} +\dots)\left(\frac{dx_m}{x_m}\right)^2\;;
\een
symmetry of (\ref{Rmln}) is thus  equivalent to relation
$$
{\rm bires}\big|_{P_m}\tr\{d_PU_m d_PU_n\}={\rm bires}\big|_{P_n}\tr\{d_PU_m d_PU_n\} \;.
$$
Let us rewrite the l.h.s. of this relation in terms of bidifferential
$W(P,Q)\equiv w(P,Q) dx_P dx_Q$ according to (\ref{dPUm}), 
taking into account the behaviour (\ref{diag}) of $W(P,Q)$ on the diagonal $P=Q$:
\ben
{\rm bires}\big|_{P_m}\tr\{d_PU_m d_PU_n\}={\rm
bires}\big|_{P_m}\stackrel{1}{\tr}\left\{\stackrel{3}{\tr}\big\{
\stackrel{3}{R}_m\stackrel{13}{w}(P,P_m)\big\}
\stackrel{2}{\tr}\big\{
\stackrel{2}{R}_n\stackrel{12}{w}(P,P_n)\big\}\right\}(dx_m(P))^2
\een
\begin{equation}
=
\stackrel{1}{\tr}\,\stackrel{2}{\tr}\left\{\stackrel{1}{R}_m
\stackrel{2}{R}_n \stackrel{12}{w}(P_m,P_n)\right\}\;.
\la{trtr1}
\end{equation}
Similarly, 
\ben
{\rm bires}\big|_{P_m}\tr\{d_PU_m d_PU_n\}=
{\tro}\,{\trd}\left\{\stackrel{2}{R}_m
\stackrel{1}{R}_n \stackrel{12}{w}(P_n,P_m)\right\}
\een
\begin{equation}
={\tro}\,{\trd}\left\{\stackrel{2}{R}_m
\stackrel{1}{R}_n \Pi\stackrel{12}{w}(P_m,P_n) \Pi\right\}
={\tro}\,{\trd}\left\{\stackrel{1}{R}_m
\stackrel{2}{R}_n \stackrel{12}{w}(P_m,P_n)\right\}\;,
\la{trtr2}
\end{equation}
coinciding with (\ref{trtr1}); the last equality in (\ref{trtr2})
follows from the symmetry property (\ref{symmetry}) of $W(P,Q)$.
$\Box$

\begin{remark}
\rm \la{gen1}
 More explicit treatment is, as
usual, possible for elliptic coverings $g=1$, when the bundle $ad_\hi$
can possess meromorphic sections with single simple pole. In this case
the additive twists $C_m^\g$ are absent, and monodromy matrices
$\hi_\g$ of the bundle $\hi$ can be chosen to be independent of
$\{{\l}_m\}$. 
In this case the above construction can be nicely rewritten in terms of elliptic r-matrix \cite{Vasil03}.

\end{remark}
\begin{remark}
\rm \la{rema}
In rank $1$ case we have fixed the bidifferential $W(P,Q)$ by the requirement that all of its
$a$-periods vanish; then $W(P,Q)$ coincides with Bergmann
kernel. This kind of normalization is not possible in higher rank due
to  non-invariance of $W(P,Q)$ under tracing along topologically non-trivial
loops on ${\L}$. However, 
 in rank 1 case there exists  another way to fix $W(P,Q)$ uniquely: one can
require that ${\rm v.p.}\iint_{\L}
W(P,Q)\overline {w_\a(Q)}=0$ for any holomorphic differential $w_\a$
(in this case $W(P,Q)$ is called the Schiffer
kernel).  These conditions  have natural higher rank analog
\cite{Fay92}. Namely, we can require that  ${\rm v.p.}\,\iint_{\L}
\stackrel{2}{\tr}\{\stackrel{12}{W}(P,Q)
\stackrel{2}{f^{\dagger}_k}(Q)\}=0$ for any  $f_k \in
H^0(ad_{\hi}\otimes K)$; due to unitarity of the matrices $\hi_\a$ the
integrand is a $(1,1)$-form on  ${\L}$. If the biresidue 
at $P=Q$ is chosen to be  the unit matrix $I\otimes I$ instead of permutation
matrix $\Pi$, our definition of $W(P,Q)$  would  give  
 the Schiffer kernel corresponding to the bundle $ad_\hi$ \cite{Fay92}.
This normalization of $W(P,Q)$ leads to the following
normalization of $d_PU_m$:  ${\rm v.p.}\iint_{\L} \tr \{d_P U_m f^\dagger_k\}=0$ for any $f_k \in
H^0(ad_{\hi}\otimes K)$.

We notice that, in contrast to Bergmann kernel in rank 1, this normalization of $W(P,Q)$
makes it non-holomorphic function of ``moduli'' ${\l}_m$; therefore, in principle, the complete
system of equations should contain also equations with respect to $\{\bar{{\l}_m}\}$ 
We shall
discuss these  aspects in more details in
further publication.
\end{remark}
\vskip0.3cm

{\bf Relationship to isomonodromy deformations in higher genus.}
The link between the genus zero systems (\ref{hie1}) and isomonodromic deformations on
Riemann sphere discussed in Sect.\ref{Reliso} can be extended to arbitrary
genus  (we outline it here for  $g\geq 2$). 
Let us briefly describe the  isomonodromic deformations on the covering ${\L}$.
Consider a stable bundle $\hi$ characterized by a set of unitary matrices $\hi_\g$
for  any $\g\in\pi_1({\L})$; consider also a divisor  $Q=Q_1+\dots +Q_L$ on
${\L}$.

Introduce a Higgs field $A(P)$ which is allowed to have simple poles
at the points $Q_1,\dots,Q_L$, i.e. $A\in H^0(ad_\hi\otimes K (Q))$;
suppose that $\tr A(P)=0$.
The higher genus analog of the linear differential
equation (\ref{Phig}) looks as follows:
\begin{equation}
d_P\Psi= A(P) \Psi\;.
\la{Phigg}
 \end{equation}
where function $\Psi$ has unit determinant and satisfies the initial condition $\Psi(P_0)=I$ at
some point $P_0\in{\L}$. 
 
In analog to genus $0$ case, function $\Psi$ has regular
singularities at the points
$Q_1,\dots,Q_L$ with some monodromy matrices $M_k$, i.e. under tracing
around $Q_k$ the function $\Psi(P)$ transforms as follows:
\ben
\Psi(P)\to\Psi(P) M_k\;.
\een
Under tracing along basic cycles of ${\L}$ the  function $\Psi$ gains
left multipliers given by the matrices $\hi^\g$; in addition, it may
gain the right multipliers $M_{a_\a}$, $M_{b_\a}$, which are analogs of monodromy matrices $M_k$: 
\ben
\Psi(P^{a_\a})= \hi_{a_\a}\Psi(P) M_{a_\a}\;,\hskip0.7cm 
\Psi(P^{b_\a})= \hi_{b_\a}\Psi(P) M_{b_\a}\;.
\een
The monodromy matrices $M_k$, $M_{a_\a}$, $M_{b_\a}$ generate a
$SL(r)$ ``monodromy'' representation of the fundamental group $\pi_1({\L}\setminus \{Q_k\})$.

If we now assume that all the monodromy matrices $M_k, M_{a_\a}, M_{b_\a}$
are independent of the branch points ${\l}_m$ and projections
$\mu_k=\pi(Q_k)$ of the regular singularities, then function $\Psi$
satisfies the 
deformation equations 
\begin{equation}
\Psi_{\mu_k}=V_k\Psi\;,
\la{defeq1}
\end{equation}
\begin{equation}
\Psi_{{\l}_m}=U_m\Psi\;,
\la{defeq2}
\end{equation}
where matrix $V_k$ has simple pole at the regular singularity $Q_k$; 
matrix $U_m$ has simple pole at the ramification point $P_m$; these
matrices transform as follows under the tracing along any $\g\in \pi_1({\L})$:
\ben
V_k(P^\g)= \hi_\g V_k(P)\hi_\g^{-1} + (\hi_\g)_{\mu_k}\hi_\g^{-1}\;,
\een
\ben
U_m(P^\g)= \hi_\g U_m(P)\hi_\g^{-1} + (\hi_\g)_{{\l}_m}\hi_\g^{-1}\;.
\een
Obviously, the part (\ref{defeq2}) of the deformation equations is
nothing but the linear system (\ref{lshr}) 
introduced above; therefore, the isomonodromic deformations in higher
genus correspond, as well as in genus zero, to a subset of solutions of
the  integrable systems (\ref{E1}), (\ref{E2}).

\section{Systems of rank 1, Darboux-Egoroff metrics
and systems of hydrodynamic type}

\subsection{Darboux-Egoroff metrics}

It turns out that each solution of the rank 1  system
(\ref{EDgenus}) defines a flat diagonal (pseudo-)metric in $\C^M$ (which gives rise to a flat diagonal
Darboux-Egoroff metric in $\R^M$ if additional reality and positivity conditions are imposed). In the sequel we shall
(in agreement with previous works on the subject) use the term
``Darboux-Egoroff metric'' for such pseudo-metrics in $\C^M$.

For diagonal (pseudo-) metric
\begin{equation}
ds^2=\sum_{m=1}^M \eta_{mm}\; d{\l}_m^2
\la{metric}
\end{equation}
the Christoffel symbols are given by
\begin{equation}
\Gamma_{mn}^k =0\;,\hskip0.7cm 
\Gamma_{nm}^n =\p_{{\l}_m}\log\sqrt{g_{nn}}\;,\hskip0.7cm m\neq n\neq k\;;
\la{Christ}
\end{equation}
they are related as follows to rotation coefficients $\beta_{mn}$:
\begin{equation}
\beta_{mn}= \f{\sqrt{g_{mm}}}{\sqrt{g_{nn}}}\Gamma_{mn}^m\equiv
\f{\p_{{\l}_n}\sqrt{\eta_{mm}}}{\sqrt{\eta_{nn}}}\;,
\hskip0.7cm m\neq n\;.
\la{rotco}
\end{equation}
The metric (\ref{metric}) is flat iff the rotation coefficients
satisfy the following equations:
\begin{equation}
\f{\p \beta_{mn}}{\p{\l}_l}=\beta_{ml}\beta_{ln}
\la{flatness1}
\end{equation}
for any distinct $l,m,n$, and each of $\beta_{mn}$ is invariant with
respect to simultaneous shifts along all $\{{\l}_k\}$:
\begin{equation}
\sum_{k=1}^M \f{\p \beta_{mn}}{\p{\l}_k}=0\;.
\la{shifts}
 \end{equation}

If in addition the rotation coefficients are symmetric, $\beta_{mn}=\beta_{nm}$, which is equivalent to relation
$\p_m g_{nn}=\p_n g_{mm}$, then there exists potential $U$ such that
\ben
g_{mm}=\f{\p U}{\p{\l}_m}
\een
and the metric (\ref{metric}) is called the Darboux-Egoroff metric.

We shall now prove that each solution of the 
system (\ref{EDgenus}) corresponds to Darboux-Egoroff metric.

\begin{theorem}
Let $f$ be an arbitrary solution of the system
(\ref{lsgenus}), and $\tauh(\{{\l}_m\})$ be the corresponding tau-function
defined by (\ref{deftaugen}) or (\ref{def1genus}). Then metric (\ref{metric}) 
with
\begin{equation}
g_{mm}= \f{\p\log\tauh}{\p{\l}_m}
\la{gmmm}
\end{equation}
is a Darboux-Egoroff metric in $\C^M$.
\end{theorem}
{\it Proof.} Let us compute the rotation coefficients of the metric (\ref{metric}). From the
definition of tau-function (\ref{deftaugen}) we have
\ben
\sqrt{\f{\p\log\tauh}{\p{\l}_m}}= \f{f_{{\l}_m}}{v_m}\;;
\een
using the expression (\ref{tausecond}) for the  second derivative of
the tau-function, we find
\begin{equation}
\beta_{mn}=\f{1}{2}\f{(\log\tauh)_{{\l}_m{\l}_n}}{\sqrt{(\log\tauh)_{{\l}_m}}\sqrt{(\log\tauh)_{{\l}_n}}}=
\f{1}{2}b(P_m,P_n)\;.
\la{gs}
\end{equation}
These functions satisfy the  equations  (\ref{flatness1}) as a
corollary of variational formulas (\ref{derglm}). 

It remains to prove that each $b(P_m,P_n)$ satisfies equations
(\ref{shifts}). We shall use invariance of the Bergmann kernel
$B(P,Q)$ with respect to biholomorphic maps. Let us consider the
branched covering ${\L}^{\e}$  which is obtained by small $\e$-shift of
all the
ramification  points $P_m$ in ${\l}$-plane i.e. the projections  of branch points
$P_m^\e$ of ${\L}^\e$ on ${\l}$-plane are equal to ${\l}_m^\e={\l}_m+\e$;
$B^\e$ is the Bergmann kernel on ${\L}^\e$.
Denote the projections of points $P$ and $Q$ on ${\l}$-plane by ${\l}$ and
$\mu$, respectively. Define the point $P^\e$ to be the point lying
on the same sheet as $P$ and having projection ${\l}+\e$ on ${\l}$-plane;
in the same way point $Q^\e$ belongs to the same sheet as $Q$ and has
projection $\mu+\e$ on ${\l}$-plane.
Since ${\L}^\e$ can be holomorphically mapped to ${\L}$ by transformation
${\l}\to{\l}+\e$ on all the sheets, we
have
\begin{equation}
B^\e(P^\e,Q^\e)=B(P,Q)\;.
\la{SSe}
\end{equation}
Assuming that $P$ belongs to a neighbourhood of the branch point $P_m$,
and $Q$ belongs to a neighbourhood of the branch point $P_n$, we can write
down the respective local parameters as $\lp_m(P)=\sqrt{{\l}-{\l}_m}$ and
$\lp_n(Q)=\sqrt{\mu-{\l}_n}$. These parameters are obviously invariant
with
respect to simultaneous $\e$-shifts of all $\{{\l}_m\}$, ${\l}$ and $\mu$:
$\lp_m^\e(P^\e)=\lp_m(P)$ and $\lp_n^\e(Q^\e)=\lp_n(Q)$.
Therefore, equality (\ref{SSe}) induces the same relation between
$b(P,Q)\equiv B(P,Q)/d\lp(P)d\lp(Q)$:
 \begin{equation}
b^\e(P^\e,Q^\e)=b(P,Q)\;.
\la{sse}
\end{equation}
Assuming now that $P=P_m$ and $Q=P_n$ and differentiating (\ref{sse})
with respect to $\e$ at $\e=0$, we come to (\ref{shifts}).

$\Box$

Of course, besides simultaneous translations, the Bergmann kernel
is invariant with respect to any other M{\"o}bius transformation of ${\l}$-plane
performed simultaneously on all the sheets of ${\L}$. We can use  this
invariance to obtain two relations, corresponding to other
one-parametric families of M\"{o}bius transformations.

\begin{proposition}
The rotation coefficients (\ref{gs}) of the Darboux-Egoroff metric
(\ref{metric}) satisfy the following relations:
\begin{equation}
\left\{\sum_{k=1}^M{\l}_k\f{\p}{\p{\l}_k}\right\}\beta_{mn} =-\beta_{mn}\;,
\la{mob1}
\end{equation}
\begin{equation}
\left\{\sum_{k=1}^M{\l}^2_k\f{\p  }{\p{\l}_k}\right\} \beta_{mn}=-({\l}_n+{\l}_m)\beta_{mn}\;,
\la{mob2}
\end{equation}
\end{proposition}
{\it Proof.} Relation (\ref{mob1}) corresponds to invariance of the
Bergmann kernel under simultaneous  dilatation on every
sheet of ${\L}$ i.e. to the transformations 
\ben
{\l}_m\to(1+\e){\l}_m\;,\hskip0.8cm
{\l}\to(1+\e){\l}\;,\hskip0.8cm \mu\to(1+\e)\mu\;.
\een
 The new feature in comparison with
the proof of relation (\ref{shifts}) is that the local parameters are
now dependent on $\e$:
\ben
\lp^\e_m(P^\e)\equiv[(1+\e){\l}-(1+\e){\l}_m]^{1/2}=(1+\e)^{1/2}\lp_m(P)\;
\een
and
\ben
\lp^\e_n(Q^\e)\equiv[(1+\e)\mu-(1+\e){\l}_n]^{1/2}=(1+\e)^{1/2}\lp_n(Q)\;;
\een
therefore, invariance $B^\e(P^\e,Q^\e)=B(P,Q)$  of the Bergmann kernel 
translates on the level of $b(P,Q)$ as follows: 
\begin{equation}
(1+\e)b^\e(P^\e,Q^\e)=b(P,Q)\;.
\la{sse1}
\end{equation}
Differentiating this relation with respect to $\e$ at $\e$=0 via the
chain rule and choosing $P=P_m$ and $Q=P_n$, we get (\ref{mob1}).

In a similar way we can deduce (\ref{mob2}) from invariance of the
Bergmann kernel with respect to the one-parametric family of
transformations 
\ben
 {\l}\to{\l}^\e=\f{\l}{1+\e{\l}}
\een
  on each sheet of ${\L}$. 
$\Box$

\vskip0.3cm
Relation (\ref{mob1}) for the rotation coefficients can be found in
\cite{Dubr94};  relation (\ref{mob2})  seems to be new.
We also notice that all  primary differentials used in
\cite{Dubr94} to construct Frobenius manifolds from Hurwitz spaces can be obtained from
solutions (\ref{solgenus}) of the  systems (\ref{EDgenus}) by appropriate specification of the contour
$\G$ and the function $h(Q)$.

\subsection{Systems of hydrodynamic type}

According to well-known results (see, for example, review \cite{Tsar90}), to each Darboux-Egoroff metric one can associate a class
of diagonal systems of hydrodynamic type
\begin{equation}
\f{\p{\l}_m}{\p x}= V_m\f{\p{\l}_m}{\p t}\;,
\la{HDS}
\end{equation}
where functions  $V_m(\{{\l}_k\})$ (the ``characteristic speeds'') are related
to Christoffel symbols (\ref{Christ}) of the metric (\ref{metric}) via system of differential equations:
\begin{equation}
\p_m V_n =\Gamma_{nm}^n (V_m- V_n)\;.
\la{VGamma}
\end{equation}
Compatibility of equations (\ref{VGamma}) is provided by equations
(\ref{flatness1}) for rotation coefficients.

Let us choose some solution of the  system (\ref{EDgenus})
parametrized by an arbitrary function $h(P)$ on contour $l$
(\ref{solgenus}).
Then the metric coefficients (\ref{gmmm}) are given by:
\begin{equation}
g_{mm}=\left\{\f{1}{2}\oint_l h(Q)B_m(Q)\right\}^2\;.
\la{metcoeffic}
\end{equation}
The Christoffel symbols of this metric  look as follows:
\begin{equation}
\Gamma_{nm}^n= \beta_{mn}\f{\oint_l h(P)B_m(P)}{\oint_l h(P)B_n(P)}\;.
\la{Chrih}
\end{equation}

Solutions of equations (\ref{VGamma}) for these Christoffel symbols
are described by the following proposition\footnote{Formula
(\ref{Vmg}) is due to T.Grava (private communication); similar
formula for the case of Whitham equations was obtained in
\cite{Krich2}.}.
\begin{proposition}
Let $h_1(P)$ be an arbitrary  H{\"o}lder-continuous and  independent of $\{{\l}_m\}$ function on contour $l$. 
Then the functions
\begin{equation}
V_m=\f{\oint_l h_1(P) B_m(P)}{\oint_l h(P) B_m(P)}
\la{Vmg}
\end{equation}
satisfy system (\ref{VGamma}) with Christoffel coefficients given by (\ref{Christ}).
\end{proposition}
The proof of this proposition is a  simple calculation based on Rauch
variational formulas for differentials $B_m$:
\begin{equation}
\f{\p}{\p{\l}_n} B_m(P) = \b_{mn} B_n(P)
\end{equation}

To construct solutions of the system (\ref{HDS}) with characteristic speeds (\ref{Vmg}) one needs to use the following
theorem by Tsarev \cite{Tsar90}:

\begin{theorem} \la{hodograph}
Let functions $V_m({\l}_1,\dots,{\l}_M)$ satisfy equations
(\ref{VGamma}). Then system of equations
\begin{equation}
\Phi_m(\{{\l}_k\})=t+ V_m(\{{\l}_k\}) x
\la{hodog}
\end{equation}
defines
implicit solution $\{{\l}_m(x,t)\}$  of the system of hydrodynamic type (\ref{HDS}),
where $\Phi_m({\l}_1,\dots,{\l}_M)$ is  an arbitrary solution of the
system of differential equations
\begin{equation}
\f{\p_n\Phi_m}{\Phi_m-\Phi_n}=\f{\p_n V_m}{V_m-V_n}
\la{PhiV}
\end{equation}
for $m,n=1,\dots,M$.
\end{theorem}

To apply the hodograph method to any of these systems we need
 to solve also the system of equations (\ref{PhiV}) for functions
$\Phi_m$.
Obviously, the solution is given by same formulas as solution of
equations (\ref{VGamma}) for functions $V_m$, but 
with another arbitrary H{\"o}lder-continuous  function $h_2(P)$ independent of
$\{{\l}_m\}$:
\begin{equation}
\Phi_m=\f{\oint_l h_2(P)B_m(P)}{\oint_l h(P)B_m(P)}\;.
\la{PhiHF}
\end{equation}
For each choice of $h_2(P)$ 
the system of equations (\ref{hodog}) defines  the implicit solution $\{{\l}_m(x,t)\}$ of the
 system of hydrodynamic type (\ref{HDS}).

We proved the following 
\begin{theorem}
Consider system of hydrodynamic type (\ref{HDS}), where velocities $V_m$
are given by formula (\ref{Vmg}) with arbitrary  H{\"o}lder-continuous and independent of
$\{{\l}_m\}$ functions $h(P)$ and $h_1(P)$ on contour $l$. 
Let $h_2(P)$ be another arbitrary and independent of
$\{{\l}_m\}$ H{\"o}lder-continuous function on contour $l$.
Then system of $M$ equations for $M$ variables $\{{\l}_m(x,t)\}_{m=1}^M$
\begin{equation}
\oint_l\{h_2(P)+ h(P)t + h_1(P)x\}B_m(P)=0\;,\hskip0.7cm m=1,\dots,M
\la{solhodog}
\end{equation}
defines implicit solution 
$\{{\l}_m(x,t)\}$ of the system (\ref{HDS}).
\end{theorem}

As before, the condition of H{\"o}lder-continuity of functions $h$,
$h_1$ and $h_2$ on contour $l$ can be relaxed. Namely, we can
substitute the contour $l$ by an arbitrary subset of ${\L}$, and define
on this subset
three arbitrary measures independent of $\{{\l}_m\}$; the only
requirement one needs to impose is commutativity of integration in
(\ref{solhodog}) with differentiation with respect to $\{{\l}_m\}$.

\section{Summary and outlook}

In this paper we propose a new class of integrable systems of partial
differential equations associated to  spaces of generic  rational maps of fixed degree.
For maps of degree two such systems give rise to the Ernst equation
from general relativity; for maps of higher degree our systems realize
the scheme of deformation of autonomous integrable systems proposed by
Burtsev, Mikhailov and Zakharov.
We introduce the
notion of the tau-function of the new  systems, and describe their relationship 
to the matrix Riemann-Hilbert problem and the Schlesinger system.

We generalize our construction to derive integrable systems associated to 
arbitrary  Hurwitz spaces $H_{g,N}$ of meromorphic
functions of degree $N$ on Riemann surfaces of genus $g\geq 2$.

When the matrix dimension  equals $1$,
our  systems are linear; they can be
solved via  scalar Riemann-Hilbert problem on the Riemann
surfaces. Each solution of such system corresponds to a flat
diagonal  metric  (Darboux-Egoroff metric), together with
corresponding systems of hydrodynamic type and their solutions.

Our results suggest the following directions of future work. 
 We expect 
our systems  for $g\geq 2$ to be natural deformations of
two-dimensional version of Hitchin systems
 proposed in \cite{Krich4}. 
The isomonodromic deformations in higher genus briefly discussed here should be
 in close relation to existing frameworks
\cite{Olshan,KorSam,Krich3}, 
as well as to non-autonomous Hitchin systems \cite{Hitchin}.
All these links should be clarified.
The applications of the new systems 
should also be studied, especially from the point of view of
their potential relationship with structures of Frobenius type.

{\bf Acknowledgments} Our work on this paper was greatly influenced by 
Andrej Nikolaevich Tyurin. 
We thank M.Bertola, E.Ferapontov, T.Grava, and V.Sokolov for
enlightening discussions. This research was supported by the
grant of  Natural Sciences and Engineering Research Council 
of Canada, grant of Fonds pour la Formation de Chercheurs et l'Aide a la Recherche de 
Quebec and the  Faculty Research Development  Program of Concordia University.


\begin{thebibliography}{99}

\bibitem{Atiyah} 
Atiyah, M.F., Vector bundles over an elliptic curve, Proc.London Math. soc. (3) {\bf 7} 414-452 (1957)


\bibitem{BelZak78} Belinskii,V.A.,  Zakharov, V.E., Integration of the
Einstein equations by the methods of inverse scattering theory and
construction of explicit multisoliton solutions, Sov.Phys.JETP {\bf
48} (1978) 985-994

\bibitem{BuMiZa} Burtsev, S.P., Zakharov, V.E.,  Mikhailov, A.V.,
The inverse problem method with a variable spectral parameter,
Teoret. Mat. Fiz. {\bf 70}  no. 3 323-341 (1987)

\bibitem{HilKur} Courant, R., Hilbert, D., {\it Methods of Mathematical
Physics} vol.2  New York: Interscience,  1962

\bibitem{Dubr1} Dubrovin, B., Hamiltonian formalizm of Whitham-type
hierarchies and topological Landau-Ginzburg models, Commun.Math.Phys.,
{\bf 145} 195-207 (1992) 

\bibitem{Dubr94} Dubrovin, B., Geometry of 2D topological field theories, in:
{\it Integrable systems and quantum groups. Proceedings,} Montecatini
Terme, 1993, pp. 120-348, Lecture Notes in Math., v.1620, Berlin: Springer,  1996

\bibitem{Dubr98} Dubrovin, B., Painlev{\'e} transendents in
two-dimensional topological field theory, in {\it The Painlev{\'e} property}
pp. 287-412,  CRM Ser. Math. Phys.  New York: Springer, 1999

\bibitem{Fay92}Fay, J.,  Kernel functions, analytic torsion, and moduli
spaces, Memoirs of the AMS,  {\bf 96} No.464, AMS  (1992)

\bibitem{Fulton}Fulton, W., Hurwitz schemes and irreducibility of
moduli of algebraic curves, Ann. of Math., {\bf 90} 542-575 (1969)

%\bibitem{GibTsa96} J. Gibbons and S. P. Tsarev, Reductions of the Benney equations, Phys. lett. A, 211 (1996) 19.

\bibitem{GibTsa99} J.Gibbons and S.P.Tsarev, Conformal maps and reductions of the Benney equations, Phys. lett. A, 258 (1999) 263.

\bibitem{Hitchin} Hitchin, N., Stable bundles and integrable systems,
Duke Math. Journ. {\bf 54} (1) 91-114 (1987)

\bibitem{MiwJim78}Jimbo, M., Miwa, M., Ueno, K., Monodromy preserving deformations of linear ordinary differential equations with rational coefficients, I, Phys. D {\bf 2}  306-352 (1981)

\bibitem{KokKorpre} A.Kokotov, D.Korotkin, Some integrable systems on
Hurwitz spaces, archive math-ph/0112051

\bibitem{Koro88}Korotkin, D., 
Finite-gap solutions of stationary axially symmetric Einstein
equations in vacuum,  Theor. Math.Phys., 
  {\bf 77} No.1 1018-1031 (1989) 

\bibitem{KorNic96} Korotkin, D., Nicolai, H., Isomonodromic
quantization of dimensionally reduced gravity, Nucl.Phys {\bf B 475}
(1996) 397-439

\bibitem{KorSam}Korotkin, D., Samtleben, H., On the quantization of
isomonodromic deformations on the torus, Int.J.Mod.Phys. {\bf A12}
2013-2030 (1997)

\bibitem{Krich1}Krichever, I.M., Method of ``averaging'' for two-dimensional
``integrable'' equations, Funct. Anal. Appl. {\bf 22} 
no.3 200-213 (1989) 

\bibitem{Krich2}Krichever, I.M.,  Spectral theory of two-dimensional periodic operators and its applications, Russ. Math. Surv.
44 (1989) 145-225.

\bibitem{Krich3} Krichever, I.M., Isomonodromy equations on algebraic
curves, canonical transformations and Whitham equations,  archive hep-th/0112096

\bibitem{Krich4} Krichever, I.M., Vector bundles and Lax equations on algebraic curves,
Commun. Math.Phys., {\bf 229} 229-269 (2002)


\bibitem{KupMan} Kupershmidt, B.A., Manin, Yu,I., Long wave equations
with a free surface. I. Conservation laws and solutions., Functional
Analysis and its applications,  {\bf 11} No.3, 31-42 (1977) 

\bibitem{Mais78} Maison, D., Are the stationary axially symmetric
Einstein equations completely integrable? Phys.Rev.Lett. {\bf 41}
(1978) 521-524

\bibitem{NarSes}Narasimhan, M.S., Seshadri, C.S., Stable and unitary
bundles on a compact Riemann surface, Ann. of Math., {\bf 82} 540-567 (1965)

\bibitem{Olshan} Levin, A.M., Olshanetsky, M.A., Hierarchies of
isomonodromic deformations and Hitchin systems, archive hep-th/9709207

\bibitem{Rauch}
Rauch, H.E., Weierstrass points, branch points, and moduli of Riemann surfaces,
Comm. Pure Appl. Math. {\bf 12} 543-560  (1959)

\bibitem{Tsar90}Tsarev, S.P.,  Geometry of hamiltonian systems of
hydrodynamic type. Generalized hodograph method. Izvestija AN USSR
Math. {\bf 54} No.5 1048-1068 (1990)

\bibitem{Vasil03} Shramchenko, V., Integrable systems related to elliptic branched coverings, to appear

\bibitem{Zver} Zverovich, E.I., 
Boundary value problems in the theory of analytic functions in H{\"o}lder classes on Riemann surfaces.
Russ.Math.Surveys {\bf 26}  no. 1  113-179  

\end{thebibliography}
\end{document}